\documentclass[sigconf]{acmart}
\usepackage{accsupp}
\usepackage{url}
\usepackage{multirow}
\usepackage{CJK}
\copyrightyear{2026}
\acmYear{2026}
\setcopyright{cc}
\setcctype{by-nc-nd}
\acmConference[CHI '26]{Proceedings of the 2026 CHI Conference on Human Factors in Computing Systems}{April 13--17, 2026}{Barcelona, Spain}
\acmBooktitle{Proceedings of the 2026 CHI Conference on Human Factors in Computing Systems (CHI '26), April 13--17, 2026, Barcelona, Spain}
\acmPrice{}
\acmDOI{10.1145/3772318.3790624}
\acmISBN{979-8-4007-2278-3/2026/04}

\begin{document}

\title[Reimagining Sign Language Technologies]{Reimagining Sign Language Technologies: Analyzing Translation Work of Chinese Deaf Online Content Creators}

\author{Xinru Tang}
\affiliation{\institution{University of California, Irvine}\city{Irvine}\state{California}\country{USA}}
\email{xinrut1@uci.edu}

\author{Anne Marie Piper}
\affiliation{\institution{University of California, Irvine}\city{Irvine}\state{California}\country{USA}}
\email{ampiper@uci.edu}



\begin{abstract}
While sign language translation systems promise to enhance deaf people's access to information and communication, they have been met with strong skepticism from deaf communities due to risks of misrepresenting and oversimplifying the richness of signed communication in technologies. This article provides empirical evidence of the complexity of translation work involved in deaf communication through interviews with 13 deaf Chinese content creators who actively produce and share sign language content on video sharing platforms with both deaf and hearing audiences. By studying this unique group of content creators, our findings highlight the nuances of sign language translation, showing how deaf creators create content with multilingualism and multiculturalism in mind, support meaning making across languages and cultures, and navigate politics involved in their translation work. Grounded in these deaf-led translation practices, we draw on the sociolinguistic concept of (trans)languaging to re-conceptualize and reimagine the design of sign language translation systems.
\end{abstract}



\maketitle
\section{Introduction}
At its 2025 I/O event, Google introduced SignGemma, a forthcoming artificial intelligence (AI) model designed to translate American Sign Language (ASL) into English text~\cite{google-gemma}. Google's advances are preceded by a long history of work on this topic. Since the 1980s, when gestural interfaces and video-based techniques began to emerge, large tech companies ~\cite{microsoft-sign-kinect, microsoft-sign, desai2023asl, nvdia-asl}, start-ups ~\cite{signapse, SignForDeaf, OmniBridge}, and research laboratories ~\cite{prietch2022systematic, starner1995real, brashear2003using, zafrulla2011american, starner2002real} have introduced numerous sign language technologies, often with the promise of automatic translation between signed language and written/spoken language. While such innovations frequently attract media attention and funding ~\cite{sign-avatar, sign-glove-ucla, sign-glove, angelini2024bridging}, sign language translation technologies have faced sharp scrutiny within deaf\footnote{We use `deaf' to refer to deaf sign language users encompassing all levels of signing literacy. Deaf communities in the U.S. often capitalize the `D' in deaf to emphasize a cultural identity. We do not differentiate between Deaf and deaf in this study, as this distinction is increasingly contested within deaf studies ~\cite{kusters2017innovations}. We use deaf throughout to acknowledge the fluidity of identity and to recognize that access to deaf cultural resources itself can be a form of privilege. We use `deaf and hard-of-hearing' (DHH) to encompass the broader population with hearing disabilities, including those who do not use sign languages. We also use DHH when our cited references use the terminology.} communities ~\cite{glove_reject, alex-lu-gloves, de2021good, de2025deaf, angelini2023contrasting, angelini2024bridging}. For example, for years, a prominent online deaf community has banned all posts proposing technology ideas intended to ``help'' deaf people, citing a fundamental lack of understanding of deaf communities as a key concern ~\cite{deaf-reddit}.

Beneath this divide lies a decades-long history of adoption and resistance within deaf communities regarding how sign languages are represented and translated through technologies. As one of the largest linguistic minorities ~\cite{mitchell2023many}, deaf sign language users have long been at the center of communication and language technology innovations, as seen in developments ranging from video relay services mediated by human interpreters ~\cite{trs-fcc, brunson2011video, skinner2018interpreting}, to automatic sign language translation systems ~\cite{bragg2019sign}, and animated signing avatars ~\cite{avatar-wfd}. Young et al. refer to `the translated deaf self' to emphasize deaf signers' ``lifelong experiences of being encountered by others and inter-subjectively known in a translated form'' ~\cite{young2019translated}. While they used the concept primarily to highlight the mediated role of interpreters in deaf-hearing communication ~\cite{young2019translated}, deaf people experience translation\footnote{In the context of spoken/written language, translation typically involves converting written text, while interpretation typically refers to converting spoken words from one language to another. We use the term translation throughout to broadly include the various forms of language and media mediation in deaf communication, both asynchronous and synchronous.} more broadly in their daily communication --- such as when they sign concepts from spoken languages and hearing cultures ~\cite{holmstrom2018deaf, hin2022translanguaging}, or sign with assistive tools like captioning ~\cite{wang2018accessibility, holmstrom2018deaf, chen2024towards}. 

The complex forms of translation experienced by deaf people often pose challenges to technology design, leading to oversimplified representations of signed communication and eliciting strong pushback from deaf communities ~\cite{glove_reject, alex-lu-gloves, de2021good, de2025deaf, angelini2023contrasting, angelini2024bridging}. A well-known example is sign language gloves. Although they marked an important advancement in sign language technology by translating gestures into English characters, they have been criticized for neglecting essential grammatical elements, such as facial expressions and body movements ~\cite{glove_reject, alex-lu-gloves}. These prevalent misrepresentations of sign languages have fostered skepticism within deaf communities towards emerging technologies that claim to support sign languages, even for technologies that support more visual and embodied interactions such as signing avatars or sign language models capable of processing visual data ~\cite{tran2023us, de2025deaf, avatar-wfd, angelini2024bridging, angelini2023contrasting}. Deaf and hard-of-hearing (DHH) scholars ~\cite{desai2023asl, de2025deaf}, and the broader DHH communities, have revealed wide concerns over sign language translation technologies, including cultural appropriation, linguistic misrepresentation, and erosion of linguistic rights ~\cite{tran2023us}. Growing concerns have been raised that the push for sign language AI could undermine deaf communities' hard-won linguistic rights if such systems become the norm ~\cite{de2025deaf, tran2023us}, especially considering that these systems are often benchmarked against human interpreters even as the quality of human interpreting is often questioned ~\cite{desai2024systemic}. Furthermore, a recent survey with 35 machine learning experts shows that misconceptions still persist even among those with sign language processing experience ~\cite{kamikubo2025exploring}. Together, this growing body of evidence points to the need for a more comprehensive understanding of sign language and its translation.

To further understand the complexities of translation involved in deaf communication and expand how sign language could be translated and represented through technologies, we turn to a growing sign language space cultivated by deaf people themselves: online sign language content created by Chinese deaf content creators ~\cite{xinru_deaf, cao2023sparkling}. We view their work as involving both content creation and translation, using the two terms together as they are inextricably linked and co-constitutive. On one hand, as content creators, they have translated a wide range of content for diverse audiences (both DHH and hearing), including news, professional knowledge, and cultural knowledge ~\cite{xinru_deaf, lu2025sound}. Yet, even when their content is not explicitly created for translation, translation often remains an integral part of their work due to the inherently cross-lingual and cross-cultural nature of deaf communication ~\cite{lu2025sound, cao2023sparkling, xinru_deaf}. These diverse and nuanced forms of translation reflect the creators' multilingual expertise and the extensive labor involved in developing a minoritized language and community-based knowledge. As such, their work offers a fertile ground for understanding sign language translation across contexts.

Drawing on interviews with 13 Chinese deaf online content creators, we uncover the complex translation work participants performed -- work that is not only linguistic, but also deeply cultural and political. Instead of turning Chinese\footnote{We use `Chinese' to refer to both the written and spoken forms, while noting that participants used Mandarin as the spoken form in their videos. Mandarin is the official and most widely spoken variety of Chinese in China, which also has a standardized writing system.} into Chinese Sign Languages (CSL) or vice versa, we observed that they mixed a range of languages and communication elements in videos to bridge diverse languages and cultures across deaf and hearing individuals. The multimodal nature of video enabled them to practice language as a living activity through signing, speech, captions, and images, weaving together this rich repertoire of linguistic, visual, and cultural resources for communication. These practices transcend traditional notions of translation, reflecting what (socio)linguists call (trans)languaging or the blending of languages and other communicative resources, thus blurring the boundaries between languages and between linguistic and non-linguistic systems \cite{love2017languaging, wei2018translanguaging, de2019describe, kusters2017beyond, henner2023unsettling}. Yet, these interlingual and cultural flexibilities also give rise to a need to navigate the politics embedded in such multiplicity of languages and cultures.

This work makes the following contributions to the HCI and accessible computing literature. First, our study extends prior accounts of sign language translation by centering translation work and practices among Chinese deaf online content creators. China presents a complex landscape for sign language communication and translation as it lacks a standardized national sign language and has diverse sign language variants ~\cite{china-sign-geography, ma2020study}. Second, our analysis offers a critical perspective on translation in deaf communication. We draw on the concept of languaging from linguistics, calling attention to the broad communication space in which sign language translation takes place, as well as the complex politics involved in navigating the multiplicity of languages and cultures. Third, we conclude with recommendations for how future work on sign language technologies can move beyond the goal of turning sign languages to written/spoken languages (or vice versa), supporting the diverse multilingual and multicultural communication practices within deaf communities, and the thriving of sign language itself.

\section{Deaf Communication and Sign Languages}
Before focusing on CSL, we briefly review the complexities of deaf communication and sign languages to provide essential context for understanding sign language and deaf communication in general. Deaf communication is a multifaceted system that is characterized by multilingualism, multimodality, and multiculturalism. Each deaf person draws on a distinct mix of languages (e.g., English and ASL), communication modes (e.g., gaze, lip-reading, gestures, body orientations and movements, and assistive tools like captioning), and cultural frameworks (e.g., expressing math concepts originally coined in English through ASL), depending on their communication partners and personal preferences ~\cite{kusters2019language, holmstrom2018deaf, wang2018accessibility, chen2024towards}. The National Deaf Center in the U.S. charactizes this diversity of deaf communication by stating that ``There's no one way to be deaf, and deaf people communicate in all kinds of ways -- both with each other and with hearing people.'' ~\cite{ndc-deaf}

A key source of complexity in deaf communication lies in sign language itself ~\cite{bragg2019sign}. Sign languages have independent vocabularies, grammars, and syntactic structures that are fundamentally distinct from written/spoken languages ~\cite{de2019legal}. Sign languages rely on visual-spatial elements, including facial expressions, body movement, and locations of signs, to convey meaning. Signers use space around their bodies and sign in non-linear structures instead of using words in linear orders as in spoken/written languages. Consequently, from a cultural perspective, deaf people often identify as part of a linguistic and cultural minority ~\cite{padden1988deaf}. 

Sign languages have rich national and regional variations. Kusters et al. noted that the naming of sign languages is inherently political; if every regional and urban variety was given a distinct name, Indonesia alone would have more than 500 named sign languages ~\cite{kusters2020sign}. Furthermore, sign language exhibits rich variation due to frequent language contact, a common phenomenon in minority languages, where interactions between different languages (including variants) lead to language switching or even new languages ~\cite{kusters2020sign}. Some linguists have thus described sign language as a ``continuum'' rather than a fixed system ~\cite{kusters2020sign}. 

Yet this richness and complexity in deaf communication has long been under-recognized. It was not until the 1960s that William Stokoe provided formal linguistic evidence that ASL is a fully  developed language~\cite{stokoe2005sign}. Before that, sign language was often dismissed as an invalid form of communication, considered a poor substitute for spoken language. In 1880 in Milan, the International Congress on Education of the Deaf's oralist proponents (i.e., people who believe that deaf education should center spoken language), voted to ban sign language~\cite{dark-age}. This event ushered in a period often referred to as the ``Dark Ages'' of deaf education~\cite{dark-age}. Besides, many signing systems taught at deaf schools or used by interpreters (often hearing) did not reflect forms naturally developed within deaf communities but enforced structures of written/spoken languages~\cite{nakamura2006deaf, gustason1980signing}. For example, Signing Exact English is a commonly used signing system that represents English using signs from ASL ~\cite{gustason1980signing}. Fundamentally, it remains English because it preserves English grammar, much like how a direct word-for-word translation from French to English would still mostly follow French structures. For example, translating ``soixante-dix'' (the French term for 70) as ``sixty-ten'' would appear unnatural to English speakers. The variety of signing systems and the minoritized position of these languages has led to a complex history for sign language, making it crucial to represent signed communication responsibly in all related initiatives.

\section{Related Work}
\subsection{Sign Language Technologies}
Sign language technologies refer to a body of tools and systems that cover sign language recognition, generation, and translation, with bidirectional transformation between signed and spoken or written languages often deemed as the ultimate goal ~\cite{bragg2019sign}. The idea of creating sign language translation machines dates back to the 1980s, when computing researchers first began exploring gestural interfaces ~\cite{glove_reject, sturman1994survey}, which gained renewed interest with the advent of video technologies in the 1990s ~\cite{starner1995real, brashear2003using, starner2002real}. Since then, research has explored systems focusing on sign language recognition (e.g., ASL dictionary search ~\cite{bragg2015user, kosa2025exploring}, ASL conversational interfaces ~\cite{glasser2020accessibility}), generation (e.g., ASL signing avatar generation ~\cite{huenerfauth2014release}), and translation (e.g., text-to-sign ~\cite{esselink2024exploring}, speech-to-sign ~\cite{cox2002tessa, glauert2006vanessa}, and sign-to-text ~\cite{camgoz2020multi}). Recently, large language models (LLMs) and multimodal LLMs have introduced new opportunities given their demonstrated capabilities in language processing tasks ~\cite{zhang2025towards, inan2025signalignlm}. For example, Zhang et al. explored LLMs and video generation models in generating ASL with non-manual markers ~\cite{zhang2025towards}. 

Despite considerable work, barriers remain to developing reliable sign language systems for real-world adoption (see ~\cite{bragg2019sign} for a review). A major bottleneck is the lack of quality sign language datasets ~\cite{bragg2021fate, bragg2019sign}. Existing datasets are limited in size, video quality, continuous signing, inclusion of native signers, and signer diversity ~\cite{bragg2019sign, zhang2025towards}. Consequently, there have been concerted efforts to collect sign language data from signing communities ~\cite{bragg2022exploring, desai2023asl, bragg2021asl, glasser2022asl, chua2025emosign, kezar2023sem}, exploring methods like interpreting Wikipedia articles ~\cite{glasser2022asl}, gamification ~\cite{bragg2021asl}, and crowdsourcing ~\cite{bragg2022exploring}. However, how to responsibly collect data from communities at scale remains an open and pressing question ~\cite{bragg2021fate}.

A deeper challenge lies in representing the expressiveness of sign language through computational forms. There is still no standardized annotation scheme for sign language data ~\cite{bragg2019sign, bragg2021fate, desai2024systemic}. Bragg et al. discussed major label formats such as Gloss\footnote{A written representation of signs using spoken/written language text. For example, ``NAME YOU ?'' corresponds to ``What is your name?'' in English.}, full translation into spoken languages, linguistic notation systems, and sign language writing systems ~\cite{bragg2021fate}. Among these, Gloss is a widely adopted approach, used either as the main output or as an intermediate representation ~\cite{desai2024systemic}. However, as Desai et al. have noted, ``glosses do not stand alone as a complete representation, and lose meaning like any translation'' ~\cite{desai2024systemic}. Given the limitations of all current representations, choosing an appropriate representation scheme requires careful consideration and design ~\cite{bragg2021fate}. For example, using reduced feature sets might be viable for specific tasks such as dictionary search ~\cite{kosa2025exploring}, while a generation model might need more sophisticated annotations for fuller representation ~\cite{zhang2025towards}. Our study seeks to contribute to this ongoing conversation about sign language representation and translation by exploring how deaf creators engage with and translate sign languages.

\subsection{Critiques of Sign Language Technologies}
With growing recognition of the risks of misrepresentation, ensuring the responsible development of sign language technologies has become a pressing concern~\cite{bragg2021fate, de2021good}. Much of the existing work highlights challenges in capturing the linguistic richness of sign languages and the potential pitfalls of translation, whether into another language or a different representational form. For example, a recent deaf-led systematic review of sign language AI research identified major issues, including the use of non-representative datasets, annotations lacking linguistic grounding, and flawed modeling approaches~\cite{desai2024systemic}. An underlying concern is that existing sign language datasets are often created without the participation of deaf stakeholders in data interpretation and quality assurance~\cite{desai2024systemic}. Therefore, these datasets may miss the embodied knowledge of disabled people that is often essential to ensure data quality~\cite{tang2026disability, garg2025s}.

Other critiques pointed to ableist assumptions about deaf communication and the resulting framings and design choices shaped by these ideologies ~\cite{desai2024systemic, de2025deaf}. Sign language technologies are frequently framed as solutions to serving deaf people when interpreters are unavailable, with sign language interpreters commonly used as the benchmark for evaluating their quality ~\cite{desai2024systemic, de2025deaf}. Desai et al. observed that research on sign language technologies is typically motivated by the goal of ``mitigating communication barriers'' for deaf people ~\cite{desai2024systemic}. These assumptions reflect the longstanding conception of interpreting as the default model for providing access, while ignoring the collaborative role that deaf people play and shared responsibilities involved in human communication ~\cite{de2021sign, de2025deaf}. Using interpreters as benchmarks also raises the question of who these technologies aim to serve, i.e., deaf people, interpreters, or their hearing communication partners ~\cite{de2021good}? Many deaf people and scholars are thus concerned that the push for sign language AI may undermine deaf communities' hard-won linguistic rights ~\cite{de2025deaf, tran2023us}, although some voices within deaf communities have also expressed that AI could offer promising alternatives to reduce the labor involved in requesting and working with human interpreters ~\cite{de2025deaf}.

\subsection{Deaf Content Creators on Video-Sharing Platforms}
The minority status of sign languages positions deaf content creators as crucial contributors to the creation and dissemination of sign language content~\cite{xinru_deaf}. Although sign language has gained legal recognition and become an integral part of accessibility and telecommunications services in many countries~\cite{de2019legal}, interpretation remains limited to specific programs, often has quality issues, and fails to accommodate the full diversity of sign languages~\cite{xinru_deaf, xiao2015chinese, wehrmeyer2015comprehension, de2019legal, japanese-yamauchi} or emergency situations~\cite{g2020science}. Consequently, much of the labor in creating sign language access falls on deaf community members, both online and offline~\cite{xinru_deaf, tran2026toward}.

Video platforms have become one such essential space~\cite{xinru_deaf, daily-moth, dpan}, where deaf people share information through a range of languages and modalities, including signing in videos, text (captions or writing in videos), speech (speaking or using AI-generated speech), and other expressive elements like music, images, and emojis ~\cite{lu2025sound, xinru_deaf, cao2024voices, mack2020social, cao2023sparkling}. Related work has revealed a vibrant online information ecosystem shaped by deaf creators, where they translate news and information for deaf audiences ~\cite{xinru_deaf}, and share deaf cultural experiences and awareness with hearing viewers ~\cite{lu2025sound, cao2024voices, cao2023sparkling}. Yet, what remains under-explored is how deaf creators develop sign language practices within these online spaces, which often reach large, diverse audiences.

Much of the HCI research on deaf content creators has been centered around social media accessibility, reporting issues including lighting, challenges in capturing full body views on-the-go, difficulties related to video uploading and downloading, and aligning
AI-generated speech to videos ~\cite{vogler2025barriers, mack2020social, cao2024voices, cao2023sparkling}. In addition to technical constraints, research has also examined the influence of social media platform dynamics over deaf creators' expressions ~\cite{lu2025sound, xinru_deaf, yoo2023understanding, cao2023sparkling}. A notable challenge reported in the literature is to share accessible content with both hearing and DHH viewers as the two groups differ in language use and communication preferences ~\cite{mack2020social, cao2023sparkling}. Related work also reveals how deaf creators face prevalent ableism on social media ~\cite{cao2023sparkling, lu2025sound, xinru_deaf}, which is often reinforced by algorithmic cultures that are biased against content from disabled users ~\cite{cao2023sparkling}. Deaf creators who target hearing audiences or share videos for financial reasons have reported pressures to conform to hearing norms, such as using AI-generated speech and simplifying signed expressions ~\cite{cao2023sparkling, lu2025sound}. 

In contrast to the typical focus on social media accessibility and algorithmic influences, what has received less attention is the rich translation work performed by deaf content creators, particularly those produced for deaf audiences (e.g.,  the Daily Moth ~\cite{daily-moth} and DPAN.TV ~\cite{dpan}). In China, sign language videos created and shared by deaf creators have become vital sources of information for deaf communities, as they use CSL in ways that are both linguistically and culturally accessible to deaf viewers ~\cite{xinru_deaf}. The growing popularity of video content by deaf creators, such as those studied in the present paper, stands in stark contrast to the limited reception of sign language interpretation in official news broadcasts within China ~\cite{interpretation_tv, xinru_deaf}. These official interpretations are frequently criticized as difficult to follow, as the interpreters (usually hearing) tend to use language misaligned with deaf communities ~\cite{xinru_deaf, interpretation_tv}. However, deaf creators' translation practices within these online deaf spaces remain largely unexplored, with only a few studies drawing attention to these spaces from deaf viewers' perspectives ~\cite{xinru_deaf}. Motivated by these community-driven practices, this study aims to contribute to a deeper understanding of sign language translation from the perspectives of deaf creators active in these spaces.

\section{Methods}
Our method involved in-depth interviews with thirteen deaf creators in China and an iterative process of data collection and analysis.

\begin{table*}[!t]
\resizebox{.8\textwidth}{!}{\begin{tabular}{cccccccc}
\hline
\textbf{P\#} &
\textbf{Age} &
\textbf{Gender} &
\begin{tabular}[c]{@{}c@{}}\textbf{Formal}\\\textbf{Education}\end{tabular} &
\begin{tabular}[c]{@{}c@{}}\textbf{Major} \textbf{Video}\\\textbf{Themes}\end{tabular} &
\begin{tabular}[c]{@{}c@{}}\textbf{Target}\\\textbf{Audience}\end{tabular}&
\begin{tabular}[c]{@{}c@{}}\textbf{Years of}\\\textbf{Sharing}\end{tabular}&\begin{tabular}[c]{@{}c@{}}\textbf{Interview}\\\textbf{Setting}\end{tabular}\\\hline

P1 &
  23 &
  M &
  \begin{tabular}[c]{@{}c@{}}Bachelor's\\Degree\end{tabular} &
    \begin{tabular}[c]{@{}c@{}}signed\\news,\\deaf\\awareness\end{tabular} &
  \begin{tabular}[c]{@{}c@{}}all, but\\mainly\\deaf\end{tabular}&
  3&
  \begin{tabular}[c]{@{}c@{}}text chat\\(written Chinese)\end{tabular}\\\hline
  
P2 &
  28 &
  \begin{tabular}[c]{@{}c@{}}Non-\\binary\end{tabular} &
  \begin{tabular}[c]{@{}c@{}}Master's\\Degree\end{tabular} &
  \begin{tabular}[c]{@{}c@{}}mental\\health\end{tabular} &
  DHH&
  0.5  &
  \begin{tabular}[c]{@{}c@{}}video\\conferencing\\(Mandarin)\end{tabular}\\\hline

P3 &
  29 &
  M &
  \begin{tabular}[c]{@{}c@{}}Bachelor's\\Degree\end{tabular} &
  \begin{tabular}[c]{@{}c@{}}signed rap\end{tabular} &
 all&
 2.5&
   \begin{tabular}[c]{@{}c@{}}text chat\\(written Chinese)\end{tabular}\\\hline
 
P4 &
  36 &
  M &
  \begin{tabular}[c]{@{}c@{}}Junior\\College\end{tabular} &
  \begin{tabular}[c]{@{}c@{}}signed news\end{tabular} &
  \begin{tabular}[c]{@{}c@{}}deaf\end{tabular}&
  1.5&
  \begin{tabular}[c]{@{}c@{}}video\\conferencing\\(sign languages)\end{tabular}\\\hline

P5 &
  26 &
  M &
  \begin{tabular}[c]{@{}c@{}}Bachelor's\\Degree\end{tabular} &
  \begin{tabular}[c]{@{}c@{}}deaf\\awareness\end{tabular} &
  Hearing&0.5 &
   \begin{tabular}[c]{@{}c@{}}text chat\\(written Chinese)\end{tabular}\\\hline
  
P6 &
  32 &
  M &
  \begin{tabular}[c]{@{}c@{}}Bachelor's\\Degree\end{tabular} &
  \begin{tabular}[c]{@{}c@{}}mime,\\deaf\\awareness\end{tabular} &
  \begin{tabular}[c]{@{}c@{}}Hearing\end{tabular}&2.5 &
   \begin{tabular}[c]{@{}c@{}}text chat\\(written Chinese)\end{tabular}\\\hline

P7 &
  42 &
  M &
  \begin{tabular}[c]{@{}c@{}}Bachelor's\\Degree\end{tabular} &
  \begin{tabular}[c]{@{}c@{}}math\end{tabular} &
  deaf&
  3.5&
   \begin{tabular}[c]{@{}c@{}}text chat\\(written Chinese)\end{tabular}\\\hline
  
P8 &
  35 &
  F &
  \begin{tabular}[c]{@{}c@{}}Bachelor's\\Degree\end{tabular} &
  \begin{tabular}[c]{@{}c@{}}deaf\\awareness,\\signed songs\end{tabular} &
  \begin{tabular}[c]{@{}c@{}}Hearing,\\sometimes\\deaf\end{tabular}&10+&
   \begin{tabular}[c]{@{}c@{}}text chat\\(written Chinese)\end{tabular}\\\hline

P9 &
  32 &
  M &
  \begin{tabular}[c]{@{}c@{}}High\\School\end{tabular} &
  \begin{tabular}[c]{@{}c@{}}deaf\\community news,\\e-commerce\end{tabular} &
  deaf&3.5&
   \begin{tabular}[c]{@{}c@{}}text chat\\(written Chinese)\end{tabular}\\\hline
  
P10 &
  26 &
  M &
  \begin{tabular}[c]{@{}c@{}}Bachelor's\\Degree\end{tabular} &
  \begin{tabular}[c]{@{}c@{}}deaf\\awareness\end{tabular} &
  Hearing&0.5 &
   \begin{tabular}[c]{@{}c@{}}text chat\\(written Chinese)\end{tabular} \\\hline

  P11 &
  31 &
  M &
  \begin{tabular}[c]{@{}c@{}}Junior\\College\end{tabular} &
  \begin{tabular}[c]{@{}c@{}}signed news,\\general\\knowledge,\\visual\\vernacular\end{tabular} &
  \begin{tabular}[c]{@{}c@{}}all, but\\mainly\\deaf\end{tabular}&
  4.5&
  \begin{tabular}[c]{@{}c@{}}video\\conferencing\\(sign languages)\end{tabular}\\\hline
  
  P12 &
   29 &
  F &
  \begin{tabular}[c]{@{}c@{}}Bachelor's\\Degree\end{tabular} &
  \begin{tabular}[c]{@{}c@{}}deaf\\awareness,\\general\\knowledge,\\signed news\\during COVID-19\end{tabular} &
  \begin{tabular}[c]{@{}c@{}}all\end{tabular} &
  2.5
  &
  \begin{tabular}[c]{@{}c@{}}phone call\\(Mandarin)\end{tabular}\\\hline

  P13 &
  38 &
  F &
  \begin{tabular}[c]{@{}c@{}}Junior\\College\end{tabular} &
  \begin{tabular}[c]{@{}c@{}}deaf\\awareness,\\signed news\\during COVID-19\end{tabular} &
  \begin{tabular}[c]{@{}c@{}}all\end{tabular} & 5+ &
   \begin{tabular}[c]{@{}c@{}}text chat\\(written Chinese)\end{tabular}\\
  \hline
\end{tabular}}
\caption{Participant demographics, shared content, and interview settings.}
\label{table::participants}
\end{table*}

\begin{table*}[!h]	
\centering
\captionsetup{justification=centering}
\begin{tabular}{cccc}
\hline
\textbf{P\#}      & \textbf{Major Platforms} & \textbf{Number of Followers} & \textbf{Number of Posts} \\ \hline
P1                   & Kuaishou        & 3.6k                & 84              \\ \hline
P2                   & WeChat          & N/A                 & 4               \\ \hline
P3                   & Kuaishou        & 1.5k                & 32              \\ \hline
\multirow{2}{*}{P4}  & Kuaishou        & 35.8k               & 135             \\
                    & WeChat          & N/A                 & 603             \\ \hline
\multirow{4}{*}{P5}  & Bilibili        & 16k                 & 82              \\
                    & Douyin          & 2.8k                & 63              \\
                    & Kuaishou        & 4.0k                & 82              \\
                    & Xiaohongshu     & 3.9k                & 127             \\ \hline
\multirow{3}{*}{P6}  & Bilibili        & 40k                 & 65              \\
                    & Douyin          & 15k                 & 167             \\
                    & Kuaishou        & 1.0k                & 33              \\ \hline
P7                   & Kuaishou        & 4.4k                & 240             \\ \hline
\multirow{3}{*}{P8}  & Bilibili        & 59k                 & 639             \\
                    & Douyin          & 15k                 & 334             \\
                    & Kuaishou        & 31.9k               & 178             \\ \hline
\multirow{2}{*}{P9}  & Bilibili        & 38k                 & 43              \\
                    & Kuaishou        & 240.5k              & 597             \\ \hline
P10                  & Bilibili        & 6.7k                & 21              \\ \hline
\multirow{2}{*}{P11} & Kuaishou        & 6.0k                & 112             \\
                    & WeChat          & N/A                 & 50              \\ \hline
\multirow{2}{*}{P12} & Kuaishou        & 2.6k                & 46              \\
                    & WeChat          & N/A                 & 258             \\ \hline
\multirow{2}{*}{P13} & Kuaishou        & 3.0k                & 127             \\
                    & WeChat          & N/A                 & 7               \\ \hline
\end{tabular}
\caption{Participants' channels. WeChat did not publicly display the number of followers.}
\label{table::videos}
\end{table*}

\subsection{Research Context}
Our study focused on Chinese Sign Language (CSL) and deaf online content creators in China. CSL is an independent language fundamentally different from written/spoken Chinese and its dialects, though some signs are influenced by Chinese characters, vocabularies from written/spoken languages (e.g., MP3), or local cultures (e.g., using a landmark building to represent a place) ~\cite{ren2024influence}. Sign language translation in China is particularly challenging, and often controversial, for two major reasons. First, there is no widely adopted standardized national sign language in China. What is called CSL is a family of regional variants, much like the dialects found with spoken languages ~\cite{china-sign-geography, ren2024influence, ma2020study}. Second, related interpreting and translation efforts have been complicated by prevalent misunderstanding and misrepresentation of CSL. A survey study with over 10,000 DHH signers in China shows over 90\% of them found sign language interpretation on television broadcasts confusing because of the heavy use of Signed Chinese ~\cite{interpretation_tv}. Similar to Signing Exact English, Signed Chinese is a manually coded system that imposes the grammar and word order of Mandarin onto signing. Tang et al. provide an example that helps illustrate the difference: a CSL sentence being [woman / hair / long / pointing (the third party) / know] with confused facial expressions, while the equivalent in Signed Chinese being [you / know / that / long / hair / woman / question mark] ~\cite{xinru_deaf}. Debates over these varied signing language systems are common in discussions of translation and language education in China ~\cite{china-sign-geography}. While some advocate for the standardization of CSL, others take pride in preserving their local sign languages ~\cite{china-sign-geography}. Similarly, while some oppose the use of Signed Chinese ~\cite{china-sign-geography, cctv_sc}, others are more open to incorporating it as part of their linguistic skills ~\cite{xinru_deaf}.

\subsection{Participants}
Thirteen deaf creators participated in this study (see Table~\ref{table::participants} for demographic details and Table~\ref{table::videos} for information about their channels). We recruited participants using purposive and snowball sampling methods, which are typically used with hard-to-reach populations ~\cite{johnston2010sampling}. We circulated a recruitment flyer and/or a written message within the lead author's online network, reaching out to both DHH people and researchers in disability-related fields for assistance in participant recruitment. Both the flyer and the message used written Chinese, as we intended to recruit content creators who can translate between CSL and Chinese and reach diverse audiences. The inclusion criteria for the study were: 1) identifying as deaf or DHH, 2) fluent in sign language(s), and 3) creating and sharing original sign language content for online audiences rather than for personal use.

Participants actively use sign languages in video formats (live and pre-recorded). While they also share text and image-based content, most content involves videos given the visual nature of sign languages. Each participant had a follower base in the thousands, with six having accounts that surpass 30,000 followers. Participants were active across multiple video sharing platforms, including Kuaishou, WeChat, Bilibili, Douyin, and Xiaohongshu. Despite nuanced differences in platform features, interface design, and focus on long or short-form video sharing, all of the platforms participants used support video sharing, which forms the basis for distributing sign language content. These platforms also include typical social media features, such as commenting, liking, and forwarding content. We present screenshots of these platforms' video interfaces in Appendix \ref{appendix::interfaces}.

Participants were fluent in both Chinese and signing and capable of translating between the two language systems. They shared sign language videos for varied reasons, including translating information for deaf communities, promoting CSL and deaf cultures with hearing audiences, or generating income through e-commerce. Despite differences in motivation, serving deaf communities or cultures remained a central goal. This shared commitment, together with their substantial follower base, provides a foundation for understanding sign language translation and communication through a deaf-centered lens. We offered all participants 350 RMB (approximately 50 USD) as compensation, with four choosing to participate voluntarily. This study was approved by the Institutional Review Board of our university.

\subsection{Data Collection}
We conducted semi-structured interviews between December 2022 and May 2023, with participants' informed consent and using their chosen methods of communication (see Table \ref{table::participants} for interview settings). We conducted all interviews one-on-one online in real-time. Most of the interviews were conducted through texting, phone, or Zoom calls in Mandarin or written Chinese as participants are fluent Chinese users. Some participants chose to use sign language, as they were most comfortable communicating with sign language. These sign language interviews were mediated by professional sign language interpreters participants recommended or in our network. The interviews lasted approximately 1 to 4 hours, with the text-based sessions generally taking longer. Participants were allowed to pause at any time, and the interviews were continued on another day until completed.

To inform the interviews, the lead author extensively observed Chinese deaf creators' channels (both our participants and others) to understand the content created and shared by deaf creators before, during, and after the interviews. These observations were primarily conducted to inform the interviews. She took these observations throughout the study, reviewing tens of accounts and hundreds of posts. This included observing and taking notes on the topics covered, the features used in the videos (e.g., captions and visual elements), and the comments left under them. Prior to each interview, she conducted closer observations of the participant's channels, such as reading their profiles and watching their videos. For participants who were regular live-streamers (P5, P6, and P9), she also observed their live-streaming rooms or recordings of previous live streams. After interviews, she returned to their channels to further contextualize the examples and practices creators described during the interviews.

We began this study with a broad interest in how deaf creators create and share sign language content online.Example questions we asked include: \textit{How did you start creating and sharing content? Can you walk me through a typical flow of how you created a video or drafted an article for sharing? Do you have any concerns about content creation and sharing? How do you engage with your viewers on the platform?} As interviews progressed, we adjusted our analytic focus and interview guide to explore their translation work as we found they all grappled with how to reach and make their content accessible to audiences that vary in linguistic and cultural backgrounds. We then added questions about translation issues, e.g., \textit{What do you think are the most challenging parts in sign language translation?}

\subsection{Data Analysis}
The lead author transcribed all interpreted conversations from interview recordings and text chat exchanges with participants into Chinese texts for analysis. Our analytic approach involved reflexive thematic analysis, which entails iterative and ongoing theme development along with data collection based on patterns of shared meaning among the data ~\cite{braun2021can}. We conducted open inductive coding of the cumulative interview data after each interview and regularly discussed the resulting themes. In the initial stages, we focused on the surface meaning of the data to familiarize ourselves with the data, e.g., identifying motivations for sharing, challenges in translation, and the elements participants incorporated in videos. At various stages of the analysis, we referred to videos posted by deaf creators (including those from our participants and others), to support our interpretation of the data. For instance, we located videos that used strategies participants had referenced in their interviews to help us better understand and explain participants' translation work. Through iterative coding and analytic memoing, we developed a deeper understanding of the translation work participants performed. All quotes included below were translated from the Chinese transcripts into English by the lead author. 

\subsection{Positionality}
This research was shaped by our backgrounds as hearing researchers based at a hearing-centered institution in the U.S. Both authors are hearing, non-signers, and have been raised in hearing cultures. The lead author is a native Mandarin speaker from Mainland China with basic knowledge of CSL and uses English as a second language. The second author is a native English speaker with no knowledge of any Chinese language or CSL. Our understanding of deafness was shaped through our engagement with disability and deaf studies and our research experiences with deaf communities in China or the U.S. To ensure our research was not solely shaped by hearing perspectives, we shared our research design, including the demographic information we planned to collect and our interview protocol, with a Chinese deaf professional in our network to get their feedback. Despite these efforts, we acknowledge that our interpretations of deafness remain shaped by our personal backgrounds and academic trainings. 

\subsection{Reflection on Translation in This Research}
This study is translational in nature. To enhance transparency, we offer a reflection on the translation within this research. There are at least three layers of translation that warrant attention. First, a significant part of the translation occurred between the participants and the researcher. To ensure effective communication: (1) most participants chose to use written or spoken Chinese; (2) before or at the start of each interview, the first author reviewed participants' online channels and engaged in brief conversations to help establish smooth communication; and (3) participants were encouraged to suggest interpreters with whom they felt comfortable. Still, the interviews mediated by interpreters may have issues common to interpreted interactions~\cite{de2021sign, skinner2018interpreting, metzger1999sign}. For example, we went back and forth to clarify word meanings, especially when interpreters or participants were unfamiliar with certain terms or when questions were phrased unclearly or too verbosely. Our hearing backgrounds and limited signing literacy may have also influenced participants' trust in us and shaped the stories and opinions they chose to share ~\cite{krawczyk2024ethics}.
Second, translation also took place within the research team. Since our team has only one member bilingual in Chinese and English, all data were translated from Chinese to English by a single researcher. Although a co-author reviewed the translated quotes and their contexts, the translation was inevitably shaped by the interpretation of a single translator for whom English is a second language. 
Third, we did intensive translation to convey our findings effectively in English academic writing and communication. Many of the examples used in this study were based on English and ASL because sign language research traditionally began with ASL~\cite{stokoe2005sign}. These examples, along with visual examples we present in our findings, are an attempt to communicate with HCI audiences who may lack a background in both sign language and Chinese. Given the complexities of translation in our research, we understand translation as an ongoing and interpretive communicative process rather than a fixed outcome.

\section{Findings}
Our analysis reveals extensive translation work involved in participants' content creation, such as translating news or professional knowledge for deaf viewers or translating sign languages to help hearing audiences learn about deaf culture. Across numerous instances of translation embedded in content creation, we found that the translation work performed by deaf creators went far beyond straightforward language matching. Instead, it involved nuanced meaning-making across languages, modalities, and semiotic systems. Crucially, the need for translation was not simply a matter of bridging sign and spoken languages and making content accessible, but arose from the broader linguistic and cultural heterogeneity across deaf and hearing individuals. Below, we describe three core aspects of the translation work involved in participants' content creation: creating with multilingual and multicultural translation in mind (Section \ref{sec::multilingual}), supporting meaning making across languages and cultures (Section \ref{sec::meaning}), and negotiating politics in translation (Section \ref{sec::politics}).

\subsection{Creating with Multilingual and Multicultural Translation in Mind}
\label{sec::multilingual}

\BeginAccSupp{method=pdfstringdef,unicode,Alt={Three screenshots of deaf content creators using different tools to communicate in videos or while streaming. Left: a person writes on a whiteboard with geometric diagrams. Middle: a person holds up a notebook with handwritten text toward the camera. Right: a person presents a small writing pad with text written on it.}}
\begin{figure}[t]
\centering
\includegraphics[width=8cm]{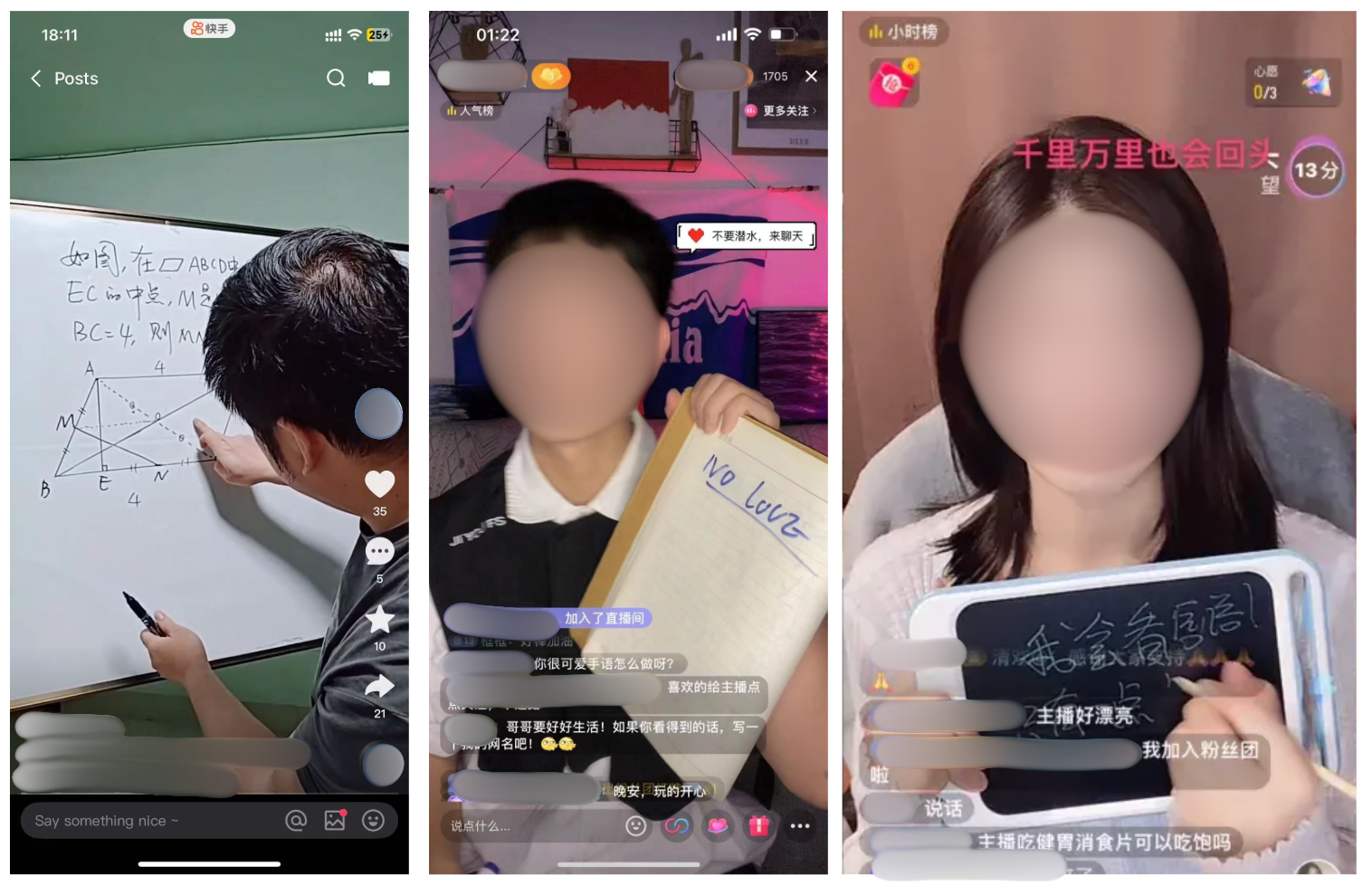}
\caption{Examples of how deaf content creators combine writing and signing using whiteboards, notebooks, and writing pads.
}
\label{fig::writing}
\end{figure}
\EndAccSupp{}

A key aspect of participants' content creation process involves engaging with the multilingualism and multiculturalism present in their audiences and content materials. Participants described encountering extensive language variation among their audiences and emphasized the need for educating and learning across cultural differences. That is, rather than translation as something they did to content after the fact, they learned to create content with multilingual and multicultural translation in mind from the start. One example of this was using signing and writing simultaneously when engaging with mixed-hearing audiences during live-streaming (see Figure \ref{fig::writing} for examples). As P5 noted:
\begin{quote}
    ``I sign while writing to ensure both [hearing and deaf viewers] are included [in live-streaming]. Deaf people often ask personal questions, which I understand, as these topics might feel more natural to them...And honestly, I still feel that hearing people aren't very familiar with deaf people.''
\end{quote}
As reflected in this quote, creators' communication with audiences involves not only language differences but also requires consideration of their diverse cultural backgrounds and knowledge systems. The mixture of language and culture P5 described stems from the diversity of people relating to DHH identities or sign languages. Besides differences between deaf and hearing viewers, others drew from personal experience to highlight the diversity within DHH populations. For instance, P1 and P3 learned Signed Chinese, rather than CSL, before high school. In contrast, while P2 was born in a deaf family, they have received oral education and did not develop a deaf identity until college where they first accessed a signing community sharing deaf pride.

The diversity of life experiences within deaf communities gives each person a unique language background, requiring participants to consider this diversity when creating content. As P2 put it,
\begin{quote}
    ``The sign language people needed was quite different from what I had imagined. Some viewers said I signed too fast. Others felt the vocabulary I used was too professional. My parents told me I didn't include enough analogies when using concepts.''
\end{quote}
The contrasts among the sign languages noted in this quote emphasize that CSL is better understood as a diverse set of language practices rather than a single, standardized language. The richness and complexity of CSL demands that even fluent deaf signers need to continually adapt and learn through lived experience.

At a deeper level, creators must bridge the distinct worldviews embedded in signed and spoken languages. This challenge is evident when P7 translates math concepts to deaf students. He explained:
\begin{quote}
    ``Hearing people are used to abstract thinking, but deaf people are more familiar with visual thinking, which is why they might leave my channels quickly. Hearing teachers would completely miss how deaf people think, as it's hard to express in oral language...Have you ever watched Tom and Jerry? That's visual thinking.''
\end{quote}

\BeginAccSupp{method=pdfstringdef,unicode,Alt={Sequence of five screenshots showing a signer using two approaches to explain the idea of turning point. (a): the person traced a angle with one hand to represent the idea of "turning"; (b) the person pointed one hand to the other, representing the idea of "point". From (c) to (d), the person traced a curve in front of body using one hand, while pointing to the bottom with another hand and sign "change" at the bottom. Green arrows highlight the motions. English translations were attached near the Chinese captions.}}
\begin{figure*}[t!]
\centering
\includegraphics[width=\linewidth]{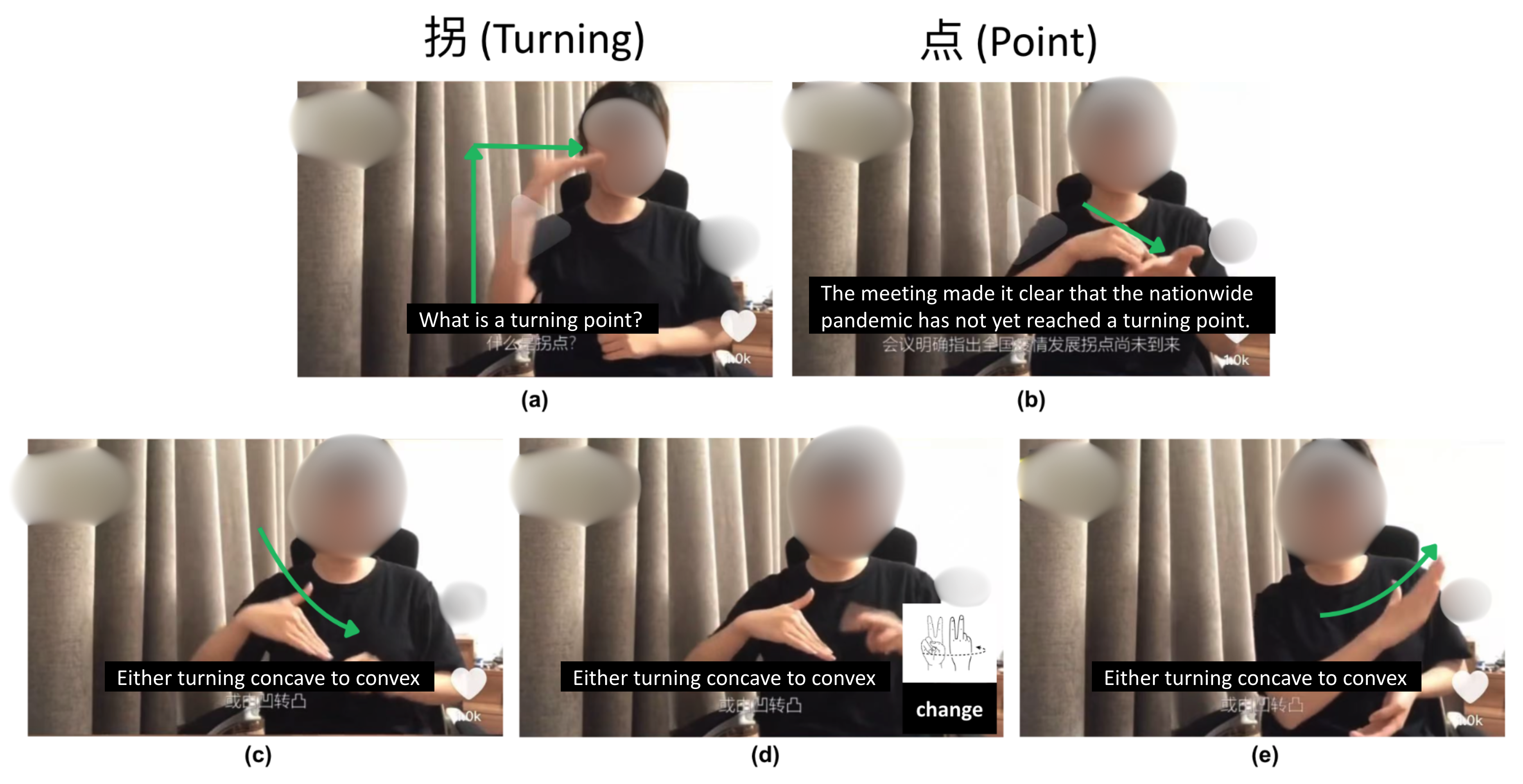}
\caption{The signer first translated the concept of `turning point' through literal mapping and then explained the concept using a visual-spatial style of signing. From (a) to (b), they used two signs to represent `turning' and `point' separately. From (c) to (d), they visually depicted a turning point. The right hand traced a curve while the other pointed downward and signed `change' to emphasize a shift or transformation at the bottom. The video is also fully captioned in Chinese. The English translations were added by the lead author.}
\label{fig::turning-point}
\end{figure*}
\EndAccSupp{}

In this case, P7 must navigate the gap between the linear structure of oral language and the non-linear, visually oriented thinking style rooted in sign language. The contrast P7 described between hearing and deaf cultures is vividly illustrated in Figure~\ref{fig::turning-point}, where a signer employed two different ways to translate the concept of `turning point' -- a mathematical term that became widely used during the COVID-19 pandemic to describe shifts in data trends. The signer began with a literal approach, using two separate signs to represent the words `turning' and `point.' Then, they transitioned into a visual-spatial style, tracing a curve with one hand and pointing to its lowest point with the other, signing `change' at that moment. This example highlights the differing expressive tendencies of written/spoken languages and signed languages: the former often relies on abstraction and conceptualization, while the latter emphasizes visual-spatial expressions that are more intuitive for deaf viewers. Given that most math materials are grounded in hearing-centered languages and cultural assumptions, translating them into a framework that resonates with deaf people remains a persistent and complex challenge. As noted by a sign language researcher in a news report, ``\textit{Only when a deaf person has a Ph.D. in physics and truly understands the field will they be able to come up with a sign to represent concepts like quantum entanglement,}'' ~\cite{cctv_sc}.

Moreover, the differences in language use often reflect deeper divergences in personal life experiences, requiring participants to navigate a range of audience preferences, knowledge backgrounds, and perspectives.  Participants discussed the challenge of gaining visibility within deaf communities, particularly when their content focused on serious topics that might lack the humor deaf viewers tend to enjoy (P1, P5, P7, \& P13).  P2 was surprised to learn that their signing style might not resonate with many deaf viewers, as it could come across as distant. She said,
\begin{quote}
    ``Some told me I looked like a well-educated person when I was signing. They may not like the style and prefer someone who's easygoing. My mother educated me that I should lower my position and status. I should practice my signing to be down-to-earth.''
\end{quote}
The feedback P2 received suggests that translating sign languages for diverse audiences requires both linguistic and cultural adaptation, with the line between these two often blurred. In contrast to P2's reception, Tang et al. reported a case in which a deaf viewer preferred content that was more in-depth and did not enjoy videos from creators with lower levels of formal education ~\cite{xinru_deaf}. These divergent views reflect the educational disparities among deaf individuals in China~\cite{lytle2005deaf}, including access to sign language education~\cite{jones2021nothing}. Research indicates that, across both K–12 and higher education, most deaf students in China have limited access to CSL instruction because the majority of teaching staff are hearing and lack formal sign language training~\cite{dingqian2019deaf, jones2021nothing}. Deaf education remains largely focused on written and spoken Chinese~\cite{dingqian2019deaf, jones2021nothing}. Besides, schools might develop different local signs~\cite{ren2024influence}. As P11 noted, ``each deaf person has a different knowledge system...Deaf education didn't become part of the national education system until the 1950s. The sign languages taught were all different.'' 

Meanwhile, those aiming to reach hearing viewers have to navigate power imbalances between deaf and hearing cultures. Lu and Guo's research shows that Chinese deaf creators often simplified sign languages into surface-level expressions due to pressure to conform to hearing norms ~\cite{lu2025sound}. A striking example participants mentioned is visual vernacular (VV), a deeply deaf art form rooted in visual storytelling ~\cite{vv-deaf}. VV combines gesture, facial expression, and body movements -- all without spoken language and reflects deeply deaf ways of communication. Despite its deep cultural significance within deaf communities, VV often struggles for recognition as a minority cultural form. P6, working as a theater actor, explained that he had never considered VV as his focus. He said,
\begin{quote}
    ``[VV] only faces deaf people. Can you [hearing people] understand it? What can you gain from it? Our target is hearing people. You need to make them pay for your show.''
\end{quote}
This statement on the limited audience for VV content, given that it ``only faces deaf people'', reflects the financial pressures and market forces that deaf creators commonly face on social media platforms and in a hearing-dominated world ~\cite{lu2025sound}. Moreover, it suggests that multilingual and multicultural translation is not just a part of their work as content creators but is \emph{the work} they are aiming to achieve.

\subsection{Supporting Meaning Making Across Languages and Cultures}
\label{sec::meaning}

A second predominant theme in participants' translation work involves how they support meaning making across the wide range of languages and cultures present among their audiences. Thus, their translation work is not simply conversion of language from one modality or representation system to another, it is about creating content such that linguistically, culturally, and educationally heterogeneous audiences can access information and develop their own understandings. To do so, participants weave together their full linguistic repertoire, leverage visual modalities that videos afford, and engage with diverse cultural frameworks to connect with their audiences. This breakdown of boundaries between communication systems lies at the core of multilingual people's language use ~\cite{wei2018translanguaging}. Consider P2, for example, who layers meaning across `straightforward' signing captured in video and `deeper' text-based captions as a way of reaching diverse viewers depending on their knowledge and literacy. They noted,
 \begin{quote}
    ``I sign in the most straightforward way, but the captions were another story. My mother might not understand the captions because they might be too deep for her. However, she could understand my signing if she hides the captions. In this way, everyone can understand my videos. People with higher Chinese literacy can read the captions. They can look at my signing if they can't.''
\end{quote}
As this quote suggests, P2 did not treat signing as a simple repetition of her captions, but mixed both to rework her content for diverse audiences. They went on to explain how they expand and reconstruct the original information through other visual and narration strategies,
\begin{quote}
    ``I would put a picture about the [concept to explain], something visual. I would do role-playing [to make deaf people understand], like what therapists would do during the sessions...If they couldn't understand `depression' or `anxiety', I would describe them through my body movements and facial expressions. `Oh, I look like nothing happened during the day, but I cry at night.' '' 
\end{quote}
What P2 described in this quote is not just to simplify languages for people with lower Chinese literacy but deep cultural work to ground the information in deaf knowledge systems and cultural norms. These efforts reflect deaf people's preferences for communication structured around visual elements ~\cite{young2016understanding}, and the intensive translation work required to adapt materials into dialogue-driven formats that deaf audiences typically prefer ~\cite{pollard2009adapting}. 

\BeginAccSupp{method=pdfstringdef,unicode,Alt={Three images of a person signing while seated indoors. In (a), both Chinese captions for 'direct bullying' and 'indirect bullying' are shown. The person pointed to the word "direct bullying". In (b), only the 'direct bullying' caption remains as the signer demonstrates with a slapping motion. In (c), the caption switches to 'indirect bullying'. English translations were attached near the Chinese captions.}}
\begin{figure*}[th]
\centering
\includegraphics[width=\linewidth]{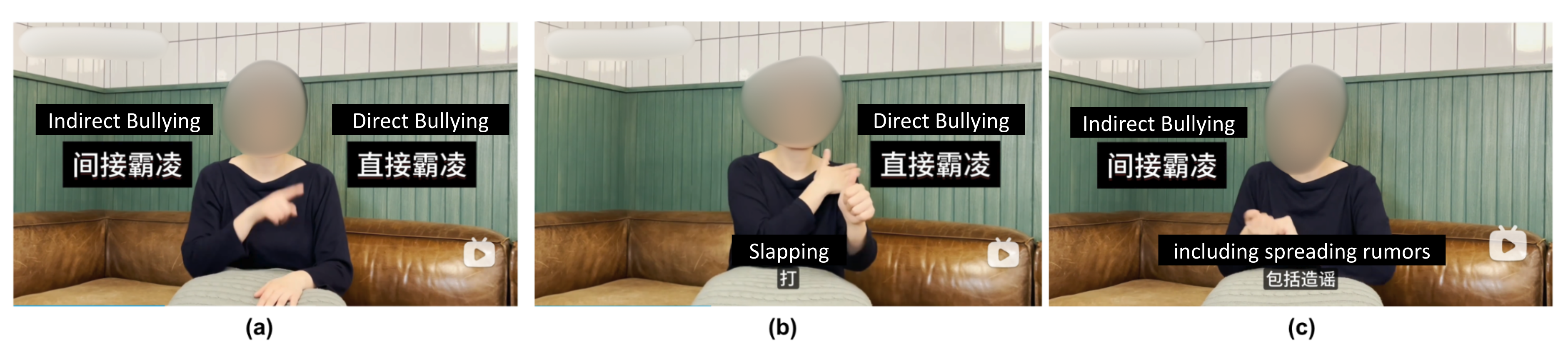}
\caption{{The signer explained the concepts of `direct bullying' and `indirect bullying' using captions and examples. (a) Both Chinese terms were displayed at the beginning. (b) The caption for `indirect bullying' disappeared as the signer turned toward the `direct bullying' caption and illustrated the concept using an example of slapping. (c) The caption for `direct bullying' disappeared and the `indirect bullying' caption reappeared as the signer explained the latter concept, using spreading rumors as an example. The video is also fully captioned in Chinese. The English translations were added by the lead author.}}
\label{fig::bully}
\end{figure*}
\EndAccSupp{}

Echoing P2's description of the wide variety of strategies they adopted to support translation, other creators described combining written, spoken, and signed languages with visuals and narration strategies to explain complex concepts originally from Chinese. In Figure~\ref{fig::bully}, for example, the signer combined captions and examples both to ground the terms they explained in their original wording (`indirect bullying' and `direct bullying') and make them visually easy to understand. The signer directed viewers' attention by controlling the appearance and disappearance of the two terms on screen and by adjusting their body orientation to clearly show which term they were explaining. They further gave examples, like kicking, slapping, and spreading rumors, to help explain the concepts. This instance provides a glimpse into the extensive support for meaning-making that deaf content creators' integrate as part of their translation work. Rather than merely mapping words from Chinese to CSL, they invested considerable effort to ensure their content was accessible and comprehensible to audiences with diverse linguistic, cultural, and educational backgrounds.

\BeginAccSupp{method=pdfstringdef,unicode,Alt={A figure with three panels labeled (a), (b), and (c), showing different examples of sign language videos from deaf content creators. Panel (a) shows a video of a person signing, with a blurred Chinese article appearing in a text box below the video. The text box is labeled "Chinese Article," and an arrow points from it to the video. Panel (b) shows a video of a person signing. A text box below the video is divided into two lines: the top line is labeled "Gloss" and contains Chinese characters, and the bottom line is labeled "Chinese Sentence" and also contains Chinese characters. An arrow points from the video to a person's shirt, which is labeled "Chinese Sentence." Panel (c) shows a video of two people signing. A text box below the video contains a Chinese sentence, and a separate text overlay on the video contains a series of Chinese words. The text box is labeled "Chinese Sentence," and the overlay is labeled "Chinese Words."}}
\begin{figure}[t]
\centering
\includegraphics[width=8cm]{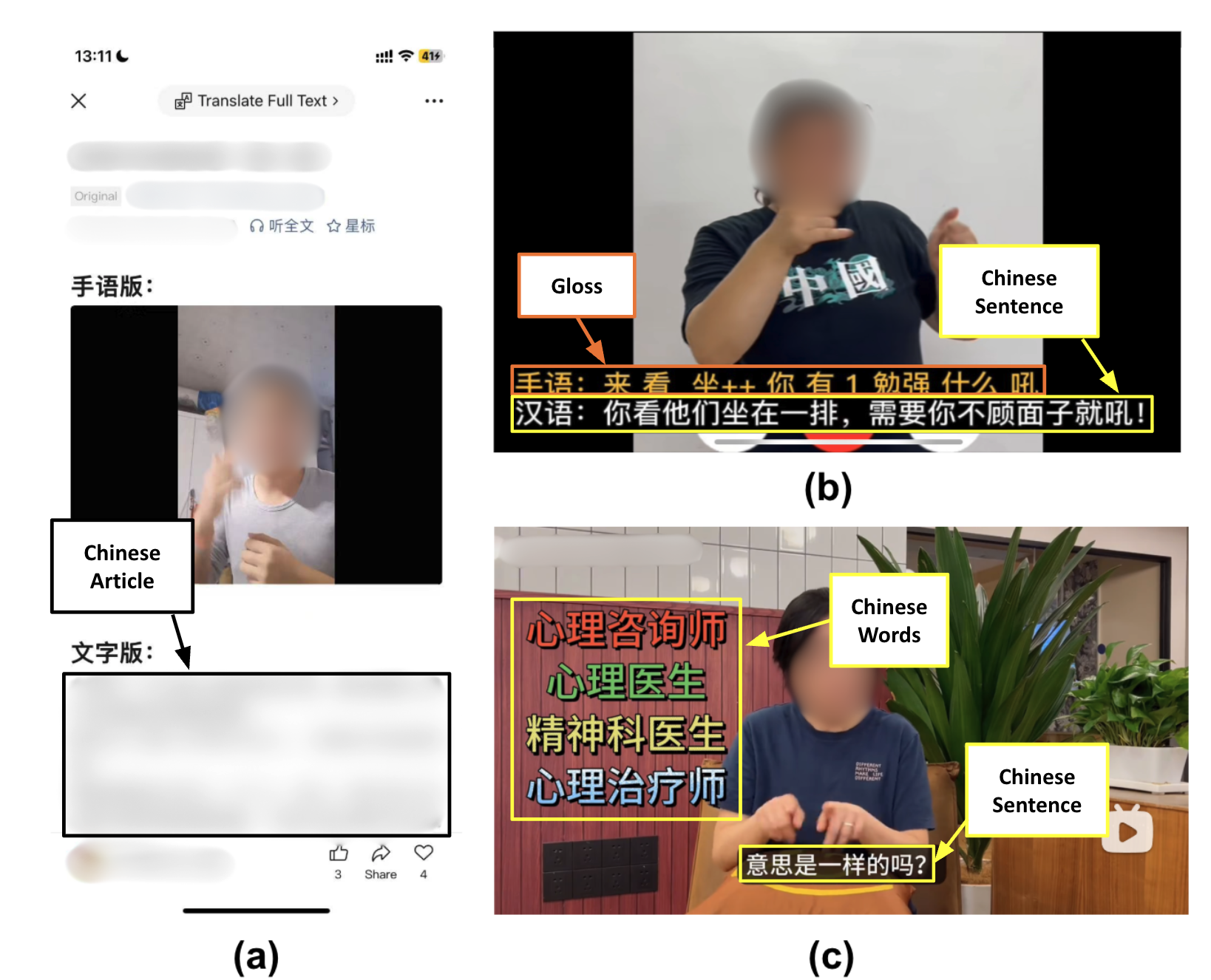}
\caption{Three examples of sign language videos by deaf content creators featuring (a) translated articles, (b) captions in gloss and Chinese full sentences, and (c) captions in Chinese words and full sentences.}
\label{fig::captions}
\end{figure}
\EndAccSupp{}

Figure~\ref{fig::captions} presents additional examples of the various language systems and captioning formats that participants mentioned or that we observed creators adopting in videos. For example, participants described gloss as being ``mainly for non-signers to recognize signs,'' because the word order of CSL differs from that of Chinese. Captioned Chinese words, meanwhile, were often used to ``help deaf viewers connect with the original Chinese concepts'' and to prevent misunderstandings in translation. Participants also incorporated other strategies alongside their signing. A common example is mouthing, during which signers silently form spoken words with their lips while signing to provide additional context or clarity ~\cite{mouthing-asl}. As P7 explained, ``signers can mouth the original Chinese characters when fingerspelling Chinese idioms (e.g., mouthing `\begin{CJK}{UTF8}{gbsn}魑 (\textcolor{purple}{Chī}) 魅 (\textcolor{purple}{Mèi}) 魍 (\textcolor{purple}{Wǎng}) 魉 (\textcolor{purple}{Liǎng})'\end{CJK} when fingerspelling `\textcolor{purple}{Ch-M-W-L}')''. A more detailed list of translation strategies mentioned by participants is provided in Appendix \ref{appendix::translation}.

\BeginAccSupp{method=pdfstringdef,unicode,Alt={A figure with four panels, labeled (a), (b), (c), and (d), showing a person signing while holding a phone. Panels (a), (b), and (c) show a person using their hands to represent the upward movement and loss of balance of a person in a "space capsule." Green arrows are superimposed on the images to indicate the direction of the hands. Panel (d) shows the person holding up their phone to display a video of a space capsule.}}
\begin{figure}[h]
\centering
\includegraphics[width=\linewidth]{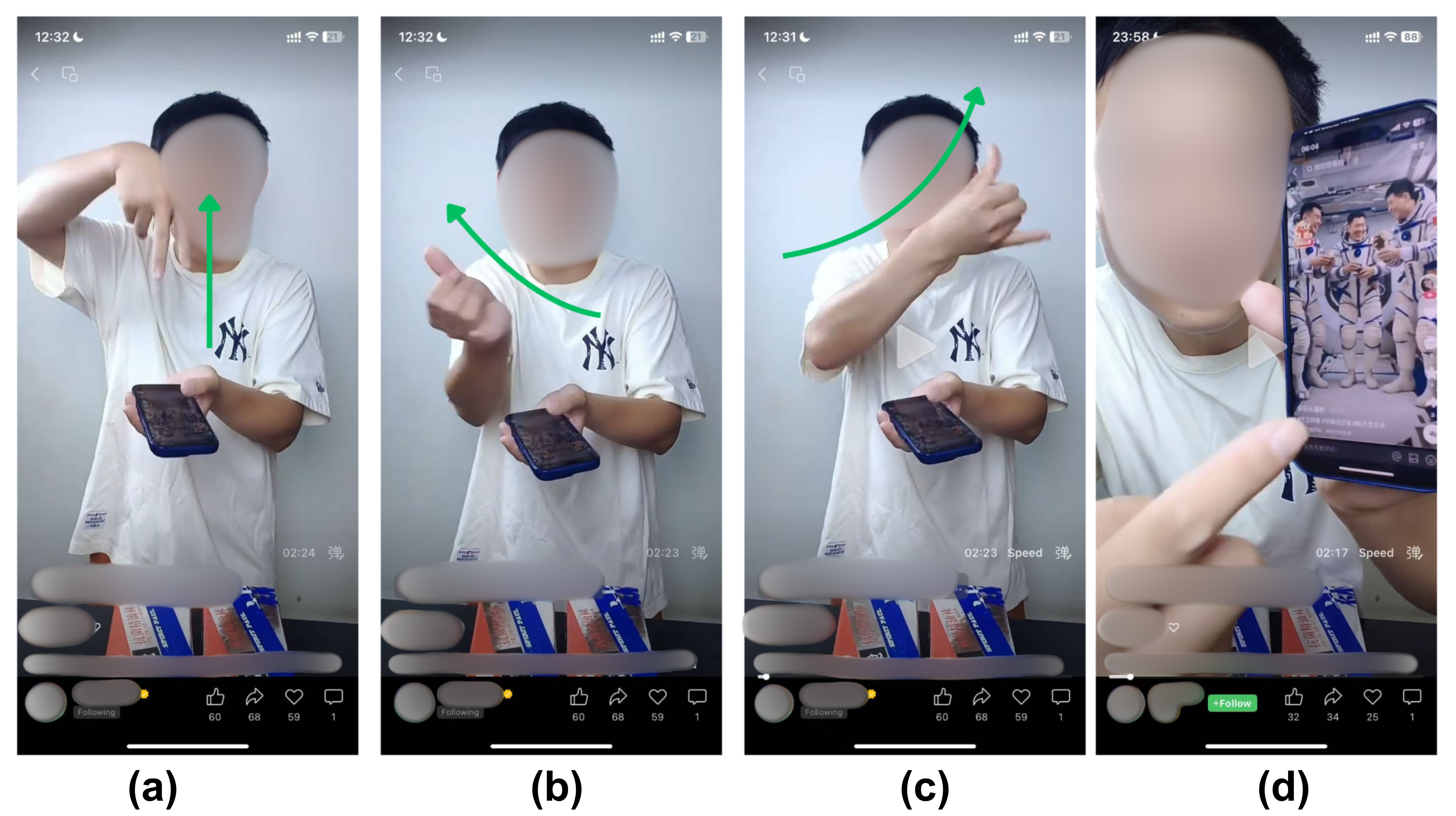}\caption{The signer combined a visual-spatial signing style with a video when talking about the concept of  `space capsule'. From (a) to (c), V-shaped (commonly used to represent legs of a person ~\cite{dengfeng-csl-classifiers}) and Y-shaped (commonly used to represent a person ~\cite{dengfeng-csl-classifiers}) classifiers were used to visually depict a person flying upward and losing balance in space. In (d), a video of a space capsule was presented to support the signing.}
\label{fig::space-capsule}
\end{figure}
\EndAccSupp{}

\BeginAccSupp{method=pdfstringdef,unicode,Alt={A figure with three panels labeled (a), (b), and (c), showing different ways to represent the word "Starbucks." Panel (a) shows the word "STARBUCKS" spelled out using the American manual alphabet. Below the letters, the phrase "Proudly served in sign language." is written. Panel (b) shows a person signing with their hands near their face, representing "curly hair." Red arrows are drawn to indicate the direction of the hand movements. Panel (c) shows the official Starbucks logo, a green and white image of a siren with long, wavy hair.}}
\begin{figure}[h]
\centering
\includegraphics[width=8cm]{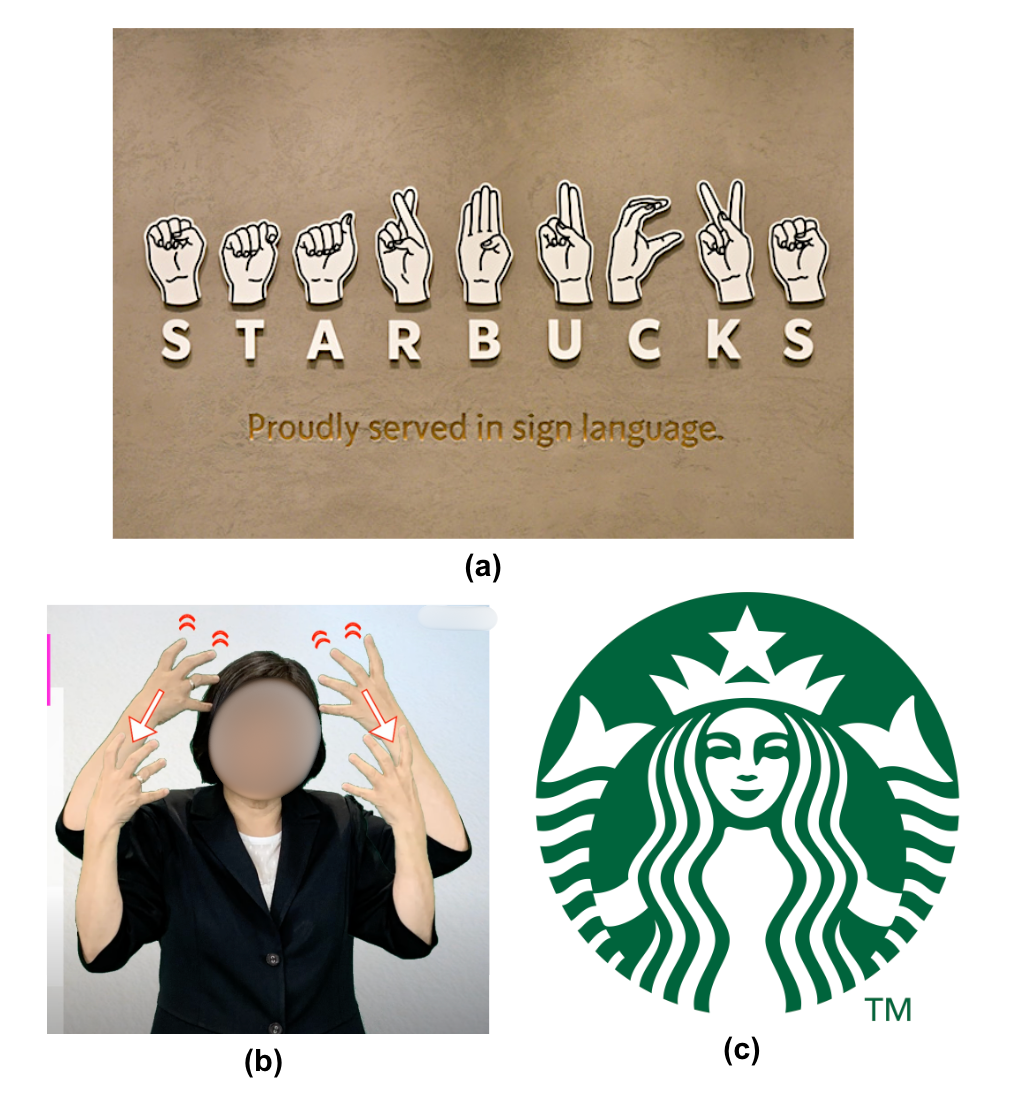}
\caption{{Multiple ways of translating `Starbucks'. (a) Fingerspelling, which accurately represents the original English word but might hard to understand for signers who do not use English (image source:~\cite{starbucks-japan}). (b) Visually-oriented translation: representing the brand icon through `curly hair'. (c) The logo of Starbucks\textregistered~ (image source: \cite{starbucks}).}} \label{fig::starbucks}
\end{figure}
\EndAccSupp{}

Across all of these examples, participants supported meaning making by bringing their intercultural understanding into their translation work. Those fluent in multiple sign language variants exhibited a flexible signing approach, readily selecting the most widely recognized signs when multiple ways of signing were present (P1 \& P4). P4 shared how he developed a blended signing style by learning from his college friends across diverse regions and incorporated it into his translation. P11 emphasized the importance of cultural understanding in translation, stating,
\begin{quote}
     ``No matter where you're from -- be it North China, South China, or anywhere in the world -- your signs will be understood as long as they align with Deaf culture. This includes using classifiers and visually depicting scenes. While there may be some differences in specific vocabulary, they won't hinder overall comprehension.''
\end{quote}
The visual style referenced in this quote is deeply rooted in sign languages and deaf culture. While many of the examples discussed in this section, such as displaying captions, are part of visual communication, a deeper approach to translation for deaf viewers involves embracing the visual-spatial style of signing, or using P7's word, ``the deaf way of thinking.'' For example, the classifiers that P11 mentioned are a type of sign unique to sign language grammar, used by signers to represent categories of nouns and to convey visual-spatial information ~\cite{classifiers-asl}. Take Figure~\ref{fig::space-capsule} as an example. To illustrate the concept of `space capsule,' the signer did not use standard signs but visually depicted a person flying upward and losing balance in space. This visual expressiveness was further enhanced by their act of incorporating a video of a space capsule. In some instances, signers may even improvise signs for translation. For example, while the signs for `Starbucks' might be different in CSL and ASL, a culturally deaf approach is to sign ``curly hair,'' to represent the brand icon (see Figure \ref{fig::starbucks}). This visually descriptive method allows people around the world, regardless of their language background, to recognize the brand, as long as they are familiar with its logo. These examples show how deaf creators navigate the variations in vocabulary across signing communities by translating in deeply `deaf' ways. Although one can translate `Starbucks' using the signs for `star' and `bucks' or by fingerspelling the word, such methods are still grounded in written or spoken language systems. In contrast, a culturally deaf approach can naturally resonate with deaf viewers regardless of their language backgrounds. This is why new signs emerge in everyday signing ~\cite{hin2022translanguaging, kusters2021emergence}, and why, at its deepest level, sign language translation should be understood not simply as transactions between languages, but as transformative and creative work that has multilingual and multicultural knowledge at its core.

\subsection{Negotiating Politics in Translation} 
\label{sec::politics}
The multiplicity of languages and cultures present in participants' translation work makes the politics of selecting and blending linguistic and communicative resources another crucial consideration. P13 cited a wide range of factors she considers in translation, explaining:
\begin{quote}
    ``It seems easy to translate, but I've put a lot of considerations into it, such as accuracy and comprehensibility, also what languages are used by people where I'm located...It's impossible to take everything into account. Otherwise, I need to make hundreds of versions."
\end{quote}
Others echoed similar points, emphasizing that creating a video involves substantial work besides technical work like editing and adding captions, especially managing the languages involved in translation. Throughout all translations, they had to adjust their approach based on their audiences' needs, the potential impact of their content, and their own identities and values.

A key tension in translation stems from its inherently transformative nature. As P8 explained, ``literal translation between CSL and Chinese is nearly impossible.'' She showed us how she translates Chinese lyrics into CSL by drawing on her understanding of context, meaning, and both languages this way,
\begin{quote}
    ``1. Looking at the whole lyrics first; 2. Breaking down the lyrics into chunks; 3. Turning to the sentence level to start the translation. Some Chinese lyrics may flip the usual word order. In these cases, [the translation] should consider how to construct the visual scene when signing.''
\end{quote}
What P8 described echoed the need to prioritize the construction of visual scenes in sign language translation ~\cite{fedorowicz2020deaf}. Signers often employ strategies such as reordering elements ~\cite{lin2021translation} or including visual details ~\cite{fedorowicz2020deaf} to prioritize visual expressiveness. The transformative nature of translation work reflects the diverse possibilities inherent in the process that might change the meaning, let alone who is able to understand it, such as decisions about when to prioritize constructing visual scenes and which signs best convey the intended meaning.

These options often gave rise to debates about the varying standards participants encountered across different contexts and the roles they assumed. For example, P11 and P13 prioritized different values in their translation approaches, emphasizing their identities as both deaf signers and translators. When we presented different translation styles for the concept of `depression' to P13, she firmly rejected blending storytelling into translation, stating,
\begin{quote}
    ``[Translating depression into finding it hard to fall asleep] is wrong. `Insomnia' is different from `depression'.''
\end{quote}
Here, P13 emphasized the distinction between the concept of depression and the associated behaviors we used to explain it. As a professional sign language interpreter, she viewed conceptual accuracy as paramount, especially when the goal was to promote knowledge. By contrast, P11 noted that accuracy sometimes conflicted with other priorities such as clarity and audience engagement. He has been promoting legal knowledge in deaf communities by explaining the Civil Code through stories he encountered or learned from other deaf people. As he put it,
\begin{quote}
    ``Translating Civil Code is supposed to be formal. However, that can be difficult for most people to fully grasp. Deaf people often ask me for clarification, so I translate it in a way that aligns with their way of thinking, although it is quite informal...However, if you're a translator or teacher, you might need to adhere to the standards.''
\end{quote}
What P11 described here not only involves challenges common to translating professional documents but also the tensions in translating hearing concepts into visually-oriented cultures embedded in sign languages. For example, he highlighted a common contradiction in sign language interpretation: to make his translation `deaf' enough, he needs to reduce the use of formal concepts and amplify facial expressions and body movements to enhance visual expressiveness. Yet, these strategies may appear exaggerated or even inappropriate in professional settings shaped by hearing cultural norms ~\cite{asl-great}. These accounts from P11 highlight the challenge of balancing established translation standards with cultural familiarity for deaf communities, particularly for someone who is not a legal professional. In navigating these tensions, P11 chose to center his own lived experience of deafness, prioritizing his goal to promote legal knowledge within deaf communities. In these two instances, P11 and P13 prioritized different values, yet both shared the goal of promoting knowledge within deaf communities. Crucially, each was keenly aware of the values they represented and the compromises they made -- an awareness shaped by their everyday experiences as deaf creators navigating a range of social and cultural contexts.

Participants also echoed the tension between authenticity and outreach in hearing-centric environments ~\cite{lu2025sound}. For example, creating videos of signed songs is a common strategy to gain visibility ~\cite{yoo2023understanding}. However, this practice has often been criticized for distorting sign languages and deaf culture, as it diverges from signing patterns within deaf communities and is often heavily influenced by spoken language structures ~\cite{maler2013songs, yoo2023understanding}. Similarly, in an effort to reach broader audiences, many participants described incorporating elements of mainstream hearing culture -- for instance, by producing videos about popular deaf-related media like the film \textit{CODA},  creating signed raps, or using Signed Chinese to help hearing viewers relate signs. While some of these efforts might be criticized as a superficial reflection of sign languages or deaf cultures, we found that these decisions were rarely made in a straightforward way. Just as participants thoughtfully navigated translation within deaf communities, those targeting hearing audiences also approached these decisions with care, balancing their goals, target audiences, and a sense of responsibility to deaf communities. For example, considering the differences between Signed Chinese and CSL, P8 combined the two language systems for his goal of promoting sign languages, stating,
\begin{quote}
    ``I decided to use CSL when teaching hearing people, while using Signed Chinese when promoting deaf or sign language awareness in general.''
\end{quote}
Similarly, P6 challenged the notion that Signed Chinese lacks value, explaining,
\begin{quote}
    ``How can you learn advanced levels without the basics? I consider Signed Chinese the basics of sign languages. Even if it's not used within deaf communities, you can treat it as a vocabulary class.''
\end{quote}
In both cases, participants did not simply choose one language over the other but rather approached their decisions with a broader perspective on their goals, even though their strategies involved compromises.

When asked about the tensions around sign language, P11 recognized the ongoing debates within deaf communities. He stressed an embodied and flexible approach, explaining,
\begin{quote}
    ``Where do you plan to use sign language? Why is the sign language used by each deaf person different? To truly understand, you need to immerse yourself in deaf communities and experience the many different ways deaf people sign. However, it's impossible to experience all the different variations because China is too big. If I were a sign language interpreter, I would focus on teaching vocabularies.'' 
\end{quote}
As referenced in this quote, the translation work involved in signed communication essentially involves navigating the complex interplay of diverse cultures, histories, and values embedded in its linguistic variations, particularly when addressing broad and heterogeneous audiences. Sustaining these efforts requires varied strategies and careful adaptation to different audiences, as participants have shown. While the compromises they make are not always without consequence, they reflect an ongoing process of learning, contestation, and negotiation that deaf people navigate as translators---either for their `deaf self' ~\cite{young2019translated}, deaf communities, or others---throughout their lives.

\section{Discussion}
Our study is motivated by the need to understand the complexity of sign language translation and human labor involved in this work, particularly as applications of AI for translation become more prevalent. In light of widespread concerns within deaf communities regarding sign language technologies, it is crucial to support deaf-led translation practices to ensure that these technologies are rooted in the linguistic and cultural knowledge of deaf communities~\cite{brace-asl}. Admittedly, the creators in this study represent a specific community engaged in broad forms of translation shaped by specific platform affordances, dynamics, and the demands of audience engagement in the context of content creation ~\cite{lu2025sound, cao2023sparkling, xinru_deaf}. However, their exposure to diverse and wide-reaching audiences provides fertile ground for them to develop and experiment with translation practices across a wide range of topics and for varied purposes. Crucially, as with all forms of translation, their translation work served to spread information and ideas, and to support human communication. Drawing on these practices developed in deaf-initiated spaces, we now revisit the conceptualization of translation and offer insights to inform the research and design of sign language translation systems.

\subsection{Towards a Languaging Approach to Sign Language Translation Technologies}
Our findings reinforce existing accounts of the challenges and complexities involved in sign language translation from professional perspectives ~\cite{bragg2019sign}. Participants echoed documented challenges in navigating structural and lexical differences between signed and spoken/written languages ~\cite{bragg2019sign}. However, their translation work extends beyond simply transferring between two language systems; it involves navigating a multiplicity of languages and cultures closely tied to both their own and their audiences' diverse identities and differential access to sign languages, such as deaf people raised with different sign language variants and knowledge systems, deaf people who learned sign languages later in life, and hearing people new to sign languages, among others. Consequently, participants engaged in translation by weaving their full linguistic repertoires, all available modalities, and other meaning-making resources, such as images, storytelling, and elements of popular culture. In doing so, they blur the boundaries between linguistic and non-linguistic elements in language use, i.e., what linguists describe as (trans)languaging ~\cite{love2017languaging, wei2018translanguaging, de2019describe, kusters2017beyond, henner2023unsettling}.

Taking a languaging perspective on translation challenges the growing conceptualization of translation as an end-to-end mapping between signs and speech or text. Rather than treating translation as a direct alignment between sign and text, participants' translation practices revealed a much broader translation space. They extended translation to the whole semiotic space that video affords and beyond, taking underlying cultural frameworks into consideration. As the idea of `languaging' suggests, language is neither established nor bounded in a single system. Rather, it is a broader activity of human beings in the world intertwined with people's socio-material environments ~\cite{goodwin2004competent, pennycook2017translanguaging} and connected to humans' ``feeling, experience, history, memory, subjectivity, and culture'' ~\cite{wei2018translanguaging}. Recognizing the full spectrum of languaging practices is therefore crucial to capturing the complexity, fluidity, and diversity inherent in human communication~\cite{henner2023unsettling, alper2018inclusive, goodwin2004competent}.

Viewing translation through the lens of languaging also questions the dominant conceptualization of translation systems limited to ``tools'' that turn sign languages into spoken/written languages ~\cite{desai2024systemic}. Participants' translation work reflects that sign language translation is not limited to mere conversion of source language or material; rather, it is a creative, interactive phenomenon emerging from deliberate acts shaped by specific purposes and the surrounding socio-cultural and political context. As our findings detail and as suppported by translation and sign language studies ~\cite{hin2022translanguaging, hodge2023deaf, lin2021translation}, signers may develop new signs, construct visual scenes, or draw on multiple languages and other communication resources for translation. This emergent, adaptive nature of signing reveals the limitations of efforts to represent signed communication within a single, standardized system. For example, Zhang et al. found that although guidelines for ASL grammar exist, signers in everyday contexts do not consistently adhere to rigid grammatical structures ~\cite{zhang2025towards}. These everyday signing and translation practices suggest that sign language translation systems should adopt a broader view of translation, treating it as an emergent communication activity rather than a task that has definitive outputs. Below we explore how researchers and designers can take action based on these insights.

\subsection{Reimagining the Design of Sign Language Translation Technologies}
Drawing on deaf people's languaging practices, we argue that future research should pursue approaches to designing sign language technologies that move beyond the dominant interpreting model (i.e., the central goal of converting between a signed language and a written/spoken language, and vice versa) and instead aim to support the diverse ways of communicating reflected within deaf communities. As Desai et al. argue, relying on the interpreting model to shape sign language technologies overlooks the unique strategies that deaf individuals have developed to navigate communication ~\cite{desai2024systemic}. Below we discuss what this means for design.

\subsubsection{Design for Sign Language Translation as a Languaging Activity}
One way to move beyond the interpreting model is to design technologies that recognize sign language as a languaging practice rather than a fixed, uniform system. A crucial step is to explore deaf people's full linguistic repertoire and communication space. Just as deaf creators blend multiple communication systems in their translation practices, sign language translation systems could adopt modular architectures that account for diverse languages, modalities, and cultural frameworks. For example, our analysis provides evidence in support of integrating multilingual captioning ~\cite{desai2025toward} and visual captioning ~\cite{liu2023visual} as part of signed interaction. Other languaging systems should also be considered such as speechreading technologies \cite{desai2013lip}.

A key to practicing sign language as a languaging activity is to acknowledge the wide linguistic variation within sign language itself. The challenges our participants encountered with vastly different signing styles provide a vivid account of deaf people's concerns about the `access hierarchy' that sign language technologies can reinforce ~\cite{de2025deaf, tran2023us} -- those who can get reliable access to education and language resources would benefit most from these technologies. While our findings show language diversity within China, many other countries face similar challenges in standardizing sign languages, such as Cambodia ~\cite{harrelson2019deaf} and Indonesia ~\cite{palfreyman2019variation}. Even in countries with an official sign language (e.g., ASL in the U.S.), considerable variation persists due to socio-cultural influences, such as racial segregation ~\cite{padden2009inside} and differences in education backgrounds ~\cite{nakamura2006deaf}. Such variations can even extend to smaller contexts such as specific schools~\cite{ren2024influence}, families ~\cite{hou2016making, hou2020signs}, and villages ~\cite{kusters2015deaf}, with many deaf people using traditional visual-spatial signing styles rather than standardized signs (see Chapter 4 of ~\cite{nakamura2006deaf} for an example). Signers also draw on what Hodge and Goswell call the nascency principle: blending languages and creating new forms of expression in response to the specific discourse and spatio-temporal context ~\cite{hodge2023deaf}, as seen in deaf content creators' daily translation practice.

To acknowledge the fluidity in sign language use, sign language technologies should be further reimagined as systems to augment deaf people's languaging practices. While efforts should be further made to enhance the representation of the diverse language variants, the hybrid and emergent nature of signed communication makes attempts to encode sign languages into translation technologies seem inevitably reductive. This is especially true given that machine translation often enforces what Ramati and Pinchevski term uniform multilingualism, i.e., processing linguistic diversity by imposing uniformity, typically through English or, more recently, through interlingual representations trained on large-scale datasets ~\cite{ramati2018uniform}. To move beyond uniformity in translation, future work can reimagine sign language technologies as extended languaging resources. For example, echoing recent explorations in this space, sign language technologies could provide multiple translation suggestions to aid human interpretation ~\cite{yin2024asl} or allow users to collaboratively discuss and refine translations ~\cite{yoo2025elmi}.

\subsubsection{Support the Thriving of Sign Language Itself}
Another crucial way to move beyond the interpreting model is to support the thriving of sign language itself. Given that in many contexts there are no established sign language standards or shared knowledge is limited, such as STEM fields ~\cite{cavender2010asl, yin2024asl}, relying on existing translation frameworks when developing sign language technologies is not sufficient. Further, as suggested in the concept of `the translated deaf self,' ~\cite{young2019translated} translation not only occurs in the device input and output but also shapes the very ontological sense of being `deaf,' pointing to the broader politics that influence deaf people's experiences with translation and their perspectives on translation technologies. For example, De Meulder cautioned against relying solely on user feedback to evaluate translation systems, noting that deaf users might lower their expectations due to a lifetime of having to tolerate inconsistent interpreting quality ~\cite{de2021good}. These complexities in how deaf people experience sign language translation highlight the need to collaborate with deaf communities, as well as professionals in domain-specific fields and translation, to support the creation of new language resources and community-driven language and translation practices. Community initiatives such as signer forums ~\cite{cavender2010asl, glasser2022asl} and deaf-centered maker spaces ~\cite{suchanek2025participation} offer promising models for such efforts.

An overemphasis on translation can also reinforce the misconception that sign languages only have meaning when rendered into another language, implying that signers' access to information and communication depends on spoken language. However, sign language is an independent, living language with its own history, culture, and linguistic richness. A key consideration that future work should center in design is the history behind sign languages and the complex language ideologies that accompany the wide spectrum of sign language use, i.e., beliefs about what constitutes `appropriate' signing and what style is `deaf' enough. Some perspectives only recognize sign languages naturally developed within deaf communities ~\cite{nakamura2006deaf}, while others take signs influenced by spoken languages ~\cite{nakamura2006deaf} as part of the broader sign language repertoire. In China, this is evident in the ongoing debate between Signed Chinese versus CSL. Signed Chinese, and other manually coded systems of written/spoken languages, have been widely criticized as `unnatural' to deaf people ~\cite{nakamura2006deaf, cctv_sc}. 

However, as our participants' translation practices demonstrate, such systems are often part of their lived linguistic repertoires, even though they might not be the most desirable options. In some cases, participants might use Signed Chinese due to broader socio-political dynamics that marginalize CSL, for example, to help hearing people map spoken language to signs, or in situations where no established CSL translation exists for concepts rooted in hearing cultures. Similarly, Signing Exact English is often taught to deaf children in English speaking countries, and many signers may incorporate such versions in their signed expression and develop new forms of languages such as Pidgin Signed English ~\cite{hauser2000code}. Equally important is that languages are constantly evolving and shaped by broader sociocultural dynamics and developments, for example in gendered expressions~\cite{burtscher2022geht, schmitz2021deaf}. Supporting explorations of these socio-political complexities requires prioritizing the growth and development of sign language itself. For example, a significant gap remains in computing systems and research dedicated to sign language linguistics ~\cite{desai2024systemic}. We hope our work sparks more deaf-led explorations in this space, since preserving the richness and histories of sign languages, and supporting their growth, depends on collective efforts grounded in the lived experiences of deaf and broader signing communities. 

\subsection{Limitations}
This study has several limitations. First, our limited proficiency in sign language may have constrained our ability to fully capture the nuances of participants' language use. Our interpretation of the videos relied largely on our limited signing literacy and the accompanying Chinese captions. Future research should involve fluent signers and incorporate multimodal analysis to more comprehensively understand signed communication. Second, our reliance on purposive and snowball sampling may have introduced bias in recruitment. Most participants were from relatively developed regions of China and had received higher education, potentially limiting the diversity of perspectives represented in this study. For example, we did not include sign languages used in ethnic minority regions, such as Korean sign language and Mongolian sign language. Future work should diversify participant characteristics, such as region, gender, age, and educational background, as individual experiences are uniquely shaped by intersecting social contexts~\cite{tang2025beyond}. Third, our analysis focused on participants' perspectives on translation, but these practices may have been shaped by platform accessibility and interface design. For example, video styles, layouts, and other affordances likely played a role in shaping deaf users' experiences on video platforms ~\cite{alkhudaidi2025perceptions}. Future research could further explore how technical affordances hinder or facilitate translation practices in digital environments. Fourth, our study did not examine audience reception, which would provide valuable insight into how translations are interpreted and co-constructed from the audience's perspective. Future research should involve a broader range of stakeholders to deepen understanding of translation.

\section{Conclusion}
This article examined the complexities of sign language translation by analyzing the translation practices among Chinese deaf online content creators. Our findings reveal the complex interplay of languages and cultures in deaf creators' translation work and in their efforts to navigate the politics embedded in the multiplicity of languages and cultures. Our study suggests that the development of sign language translation systems must include a more expansive understanding of translation, moving beyond the interpreting model to support deaf communication as a multilingual, multimodal, and multicultural activity, while also fostering the growth of sign languages themselves.

\begin{acks}
We thank the anonymous reviewers, Stacy Branham, Paul Dourish, Gillian Hayes, and members of ARC at UCI for their valuable feedback on earlier drafts of this work. We also thank our participants for sharing their experiences, and the interpreters for their support through sign language interpretation during the interviews as well as for their thoughtful feedback on our study.
\end{acks}

\bibliographystyle{ACM-Reference-Format}
\bibliography{sample-base}

@misc{alex-lu-gloves,
  author    = {Lu, Alex},
  title     = {{Deaf People Don't Need New Communication Tools -- Everyone Else Does}},
url      =  {https://medium.com/@alexijie/deaf-people-dont-need-new-communication-tools-everyone-else-does-df83b5eb28e7},
  note      = {Retrieved November 6, 2025}}

@Inbook{angelini2024bridging,
author= {Angelini, Robin and Spiel, Katta and de Meulder, Maartje},
editor= {Way, Andy and Leeson, Lorraine and Shterionov, Dimitar},
title= {{Bridging the Gap: Understanding the Intersection of Deaf and Technical Perspectives on Signing Avatars}},
bookTitle= {{Sign Language Machine Translation}},
year= {2024},
publisher= {Springer Nature Switzerland},
address= {Cham},
pages= {291--308},
isbn= {978-3-031-47362-3},
doi= {10.1007/978-3-031-47362-3_12},
url= {https://doi.org/10.1007/978-3-031-47362-3_12}
}

@inproceedings{angelini2023contrasting,
author = {Angelini, Robin},
title = {{Contrasting Technologists' and Activists' Positions on Signing Avatars}},
year = {2023},
isbn = {9781450394222},
publisher = {Association for Computing Machinery},
address = {New York, NY, USA},
url = {https://doi.org/10.1145/3544549.3583946},
doi = {10.1145/3544549.3583946},
booktitle = {Extended Abstracts of the 2023 CHI Conference on Human Factors in Computing Systems},
articleno = {566},
numpages = {6},
location = {Hamburg, Germany},
series = {CHI EA '23}
}

@inproceedings{alkhudaidi2025perceptions,
author = {Alkhudaidi, Khulood and Burke, Tish and Boll, Rachel and Mahajan, Shruti and Solovey, Erin T. and Reis, Jeanne},
title = {{Perceptions and Preferences: Deaf ASL-Signing Users' Insights on Video Elements, Styles and Layouts}},
year = {2025},
isbn = {9798400713941},
publisher = {Association for Computing Machinery},
address = {New York, NY, USA},
url = {https://doi.org/10.1145/3706598.3714296},
doi = {10.1145/3706598.3714296},
booktitle = {Proceedings of the 2025 CHI Conference on Human Factors in Computing Systems},
articleno = {65},
numpages = {20},
location = {Yokohama, Japan},
series = {CHI '25}
}

@article{alper2018inclusive,
  title={{Inclusive sensory ethnography: Studying new media and neurodiversity in everyday life}},
  author={Alper, Meryl},
  journal={New Media \& Society},
  volume={20},
  number={10},
  pages={3560--3579},
  year={2018},
  url = {https://doi.org/10.1177/1461444818755394},
  publisher={SAGE Publications Sage UK: London, England}
}

@misc{avatar-wfd,
  author    = {{World Federation of the Deaf and World Association of Sign Language Interpreters}},
  title     = {{WFD and WASLI Issue Statement on Signing Avatars}},
  url      = {https://wfdeaf.org/wfd-wasli-issue-statement-signing-avatars/},
year = {2018},
  note      = {Retrieved June 6, 2025}}

@book{brunson2011video,
  title={{Video relay service interpreters: Intricacies of sign language access}},
  author={Brunson, Jeremy L},
  year={2011},
  url = {https://doi.org/10.2307/j.ctv2rh27qm},
  publisher={Gallaudet University Press}
}

@inproceedings{burtscher2022geht,
author = {Burtscher, Sabrina and Spiel, Katta and Klausner, Lukas Daniel and Lardelli, Manuel and Gromann, Dagmar},
title = {``Es geht um Respekt, nicht um Technologie'': Erkenntnisse aus einem Interessensgruppen-\"{u}bergreifenden Workshop zu genderfairer Sprache und Sprachtechnologie},
year = {2022},
isbn = {9781450396905},
publisher = {Association for Computing Machinery},
address = {New York, NY, USA},
url = {https://doi.org/10.1145/3543758.3544213},
doi = {10.1145/3543758.3544213},
booktitle = {Proceedings of Mensch Und Computer 2022},
pages = {106-118},
numpages = {13},
location = {Darmstadt, Germany},
series = {MuC '22}
}

@INPROCEEDINGS{brashear2003using,
  author={Brashear, H. and Starner, T. and Lukowicz, P. and Junker, H.},
  booktitle={Seventh IEEE International Symposium on Wearable Computers, 2003. Proceedings.}, 
  title={{Using multiple sensors for mobile sign language recognition}}, 
  year={2003},
  volume={},
  number={},
  pages={45-52},
  doi={10.1109/ISWC.2003.1241392}}

@inproceedings{cavender2010asl,
author = {Cavender, Anna C. and Otero, Daniel S. and Bigham, Jeffrey P. and Ladner, Richard E.},
title = {{Asl-stem forum: enabling sign language to grow through online collaboration}},
year = {2010},
isbn = {9781605589299},
publisher = {Association for Computing Machinery},
address = {New York, NY, USA},
url = {https://doi.org/10.1145/1753326.1753642},
doi = {10.1145/1753326.1753642},
booktitle = {Proceedings of the SIGCHI Conference on Human Factors in Computing Systems},
pages = {2075-2078},
numpages = {4},
location = {Atlanta, Georgia, USA},
series = {CHI '10}
}

@inproceedings{bragg2015user,
author = {Bragg, Danielle and Rector, Kyle and Ladner, Richard E.},
title = {{A User-Powered American Sign Language Dictionary}},
year = {2015},
isbn = {9781450329224},
publisher = {Association for Computing Machinery},
address = {New York, NY, USA},
url = {https://doi.org/10.1145/2675133.2675226},
doi = {10.1145/2675133.2675226},
booktitle = {Proceedings of the 18th ACM Conference on Computer Supported Cooperative Work \& Social Computing},
pages = {1837-1848},
numpages = {12},
location = {Vancouver, BC, Canada},
series = {CSCW '15}
}

@article{bragg2022exploring,
author = {Bragg, Danielle and Glasser, Abraham and Minakov, Fyodor and Caselli, Naomi and Thies, William},
title = {{Exploring Collection of Sign Language Videos through Crowdsourcing}},
year = {2022},
issue_date = {November 2022},
publisher = {Association for Computing Machinery},
address = {New York, NY, USA},
volume = {6},
number = {CSCW2},
url = {https://doi.org/10.1145/3555627},
doi = {10.1145/3555627},
journal = {Proc. ACM Hum.-Comput. Interact.},
month = nov,
articleno = {514},
numpages = {24}
}

@inproceedings{bragg2021asl,
author = {Bragg, Danielle and Caselli, Naomi and Gallagher, John W. and Goldberg, Miriam and Oka, Courtney J. and Thies, William},
title = {ASL Sea Battle: Gamifying Sign Language Data Collection},
year = {2021},
isbn = {9781450380966},
publisher = {Association for Computing Machinery},
address = {New York, NY, USA},
url = {https://doi.org/10.1145/3411764.3445416},
doi = {10.1145/3411764.3445416},
booktitle = {Proceedings of the 2021 CHI Conference on Human Factors in Computing Systems},
articleno = {271},
numpages = {13},
location = {Yokohama, Japan},
series = {CHI '21}
}

@article{bragg2021fate,
author = {Bragg, Danielle and Caselli, Naomi and Hochgesang, Julie A. and Huenerfauth, Matt and Katz-Hernandez, Leah and Koller, Oscar and Kushalnagar, Raja and Vogler, Christian and Ladner, Richard E.},
title = {{The FATE Landscape of Sign Language AI Datasets: An Interdisciplinary Perspective}},
year = {2021},
issue_date = {June 2021},
publisher = {Association for Computing Machinery},
address = {New York, NY, USA},
volume = {14},
number = {2},
issn = {1936-7228},
url = {https://doi.org/10.1145/3436996},
doi = {10.1145/3436996},
journal = {ACM Trans. Access. Comput.},
month = jul,
articleno = {7},
numpages = {45},
}

@inproceedings{bragg2019sign,
author = {Bragg, Danielle and Koller, Oscar and Bellard, Mary and Berke, Larwan and Boudreault, Patrick and Braffort, Annelies and Caselli, Naomi and Huenerfauth, Matt and Kacorri, Hernisa and Verhoef, Tessa and Vogler, Christian and Ringel Morris, Meredith},
title = {{Sign Language Recognition, Generation, and Translation: An Interdisciplinary Perspective}},
year = {2019},
isbn = {9781450366762},
publisher = {Association for Computing Machinery},
address = {New York, NY, USA},
url = {https://doi.org/10.1145/3308561.3353774},
doi = {10.1145/3308561.3353774},
booktitle = {Proceedings of the 21st International ACM SIGACCESS Conference on Computers and Accessibility},
pages = {16-31},
numpages = {16},
location = {Pittsburgh, PA, USA},
series = {ASSETS '19}
}

@article{braun2021can,
  title={{Can I use TA? Should I use TA? Should I not use TA? Comparing reflexive thematic analysis and other pattern-based qualitative analytic approaches}},
  author={Braun, Virginia and Clarke, Victoria},
  journal={Counselling and Psychotherapy Research},
  volume={21},
  number={1},
  pages={37--47},
  year={2021},
  url = {https://doi.org/10.1002/capr.12360},
  publisher={Wiley Online Library}
}

@inproceedings{chen2024towards,
author = {Chen, Si and Waller, James and Seita, Matthew and Vogler, Christian and Kushalnagar, Raja and Wang, Qi},
title = {Towards Co-Creating Access and Inclusion: A Group Autoethnography on a Hearing Individual's Journey Towards Effective Communication in Mixed-Hearing Ability Higher Education Settings},
year = {2024},
isbn = {9798400703300},
publisher = {Association for Computing Machinery},
address = {New York, NY, USA},
url = {https://doi.org/10.1145/3613904.3642017},
doi = {10.1145/3613904.3642017},
booktitle = {Proceedings of the 2024 CHI Conference on Human Factors in Computing Systems},
articleno = {55},
numpages = {14},
location = {Honolulu, HI, USA},
series = {CHI '24}
}

@InProceedings{camgoz2020multi,
author= {Camgoz, Necati Cihan and Koller, Oscar and Hadfield, Simon and Bowden, Richard},
editor= {Bartoli, Adrien and Fusiello, Andrea},
title= {{Multi-channel Transformers for Multi-articulatory Sign Language Translation}},
booktitle= {Computer Vision -- ECCV 2020 Workshops},
year= {2020},
publisher= {Springer International Publishing},
address= {Cham},
pages= {301--319},
isbn= {978-3-030-66823-5},
url = {https://doi.org/10.1007/978-3-030-66823-5_18}
}

@inproceedings{cox2002tessa,
author = {Cox, Stephen and Lincoln, Michael and Tryggvason, Judy and Nakisa, Melanie and Wells, Mark and Tutt, Marcus and Abbott, Sanja},
title = {{Tessa, a system to aid communication with deaf people}},
year = {2002},
isbn = {1581134649},
publisher = {Association for Computing Machinery},
address = {New York, NY, USA},
url = {https://doi.org/10.1145/638249.638287},
doi = {10.1145/638249.638287},
booktitle = {Proceedings of the Fifth International ACM Conference on Assistive Technologies},
pages = {205-212},
numpages = {8},
location = {Edinburgh, Scotland},
series = {ASSETS '02}
}

@article{chua2025emosign,
  title={{EmoSign: A Multimodal Dataset for Understanding Emotions in American Sign Language}},
  author={Chua, Phoebe and Fang, Cathy Mengying and Ohkawa, Takehiko and Kushalnagar, Raja and Nanayakkara, Suranga and Maes, Pattie},
  journal={arXiv preprint arXiv:2505.17090},
  url = {https://arxiv.org/abs/2505.17090},
  year={2025}
}

@inproceedings{cao2023sparkling,
author = {Cao, Beiyan and He, Changyang and Zhou, Muzhi and Fan, Mingming},
title = {{Sparkling Silence: Practices and Challenges of Livestreaming Among Deaf or Hard of Hearing Streamers}},
year = {2023},
isbn = {9781450394215},
publisher = {Association for Computing Machinery},
address = {New York, NY, USA},
url = {https://doi.org/10.1145/3544548.3581053},
doi = {10.1145/3544548.3581053},
booktitle = {Proceedings of the 2023 CHI Conference on Human Factors in Computing Systems},
articleno = {58},
numpages = {15},
location = {Hamburg, Germany},
series = {CHI '23}
}

@inproceedings{cao2024voices,
author = {Cao, Jiaxun and Peng, Xuening and Liang, Fan and Tong, Xin},
title = {{``Voices Help Correlate Signs and Words'': Analyzing Deaf and Hard-of-Hearing (DHH) TikTokers' Content, Practices, and Pitfalls}},
year = {2024},
isbn = {9798400703300},
publisher = {Association for Computing Machinery},
address = {New York, NY, USA},
url = {https://doi.org/10.1145/3613904.3642413},
doi = {10.1145/3613904.3642413},
booktitle = {Proceedings of the 2024 CHI Conference on Human Factors in Computing Systems},
articleno = {34},
numpages = {18},
location = {Honolulu, HI, USA},
series = {CHI '24}
}

@misc{china-sign-geography,
  author    = {Li, Xin and Luo, Yahan},
  title     = {{Signs of Unity: Can China's Deaf Community Find a Common Language?}},
year = {2024},
month = {October},
day = {28},
url      = {https://www.sixthtone.com/news/1016095}}

@inproceedings{desai2013lip,
author = {Desai, Aashaka and Mankoff, Jennifer and Ladner, Richard E.},
title = {Understanding and Enhancing The Role of Speechreading in Online d/DHH Communication Accessibility},
year = {2023},
isbn = {9781450394215},
publisher = {Association for Computing Machinery},
address = {New York, NY, USA},
url = {https://doi.org/10.1145/3544548.3580810},
doi = {10.1145/3544548.3580810},
booktitle = {Proceedings of the 2023 CHI Conference on Human Factors in Computing Systems},
articleno = {608},
numpages = {17},
location = {Hamburg, Germany},
series = {CHI '23}
}

@article{dingqian2019deaf,
  title={Deaf education and the use of sign language in Mainland China},
  author={Dingqian, Gu and Ying, Liu and Xirong, He},
  journal={Deaf education beyond the western world: context, challenges, and prospects},
  pages={285},
  year={2019},
  url={https://doi.org/https://doi.org/10.1093/oso/9780190880514.001.0001},
  publisher={Oxford University Press}
}

@article{de2019describe,
  title={Describe, don't prescribe. The practice and politics of translanguaging in the context of deaf signers},
  author={De Meulder, Maartje and Kusters, Annelies and Moriarty, Erin and Murray, Joseph J},
  journal={Journal of Multilingual and Multicultural Development},
  volume={40},
  number={10},
  pages={892--906},
  year={2019},
  url = {https://doi.org/10.1080/01434632.2019.1592181},
  publisher={Taylor \& Francis}
}

@misc{dpan,
  author = {DPAN},
  title = {{DPAN.TV}},
  url = {https://dpan.tv/},
  year = {Retrieved February, 2023}
  }

@inproceedings{desai2023asl,
author = {Desai, Aashaka and Berger, Lauren and Minakov, Fyodor O. and Milan, Vanessa and Singh, Chinmay and Pumphrey, Kriston and Ladner, Richard E. and Daum\'{e}, Hal and Lu, Alex X. and Caselli, Naomi and Bragg, Danielle},
title = {{ASL citizen: a community-sourced dataset for advancing isolated sign language recognition}},
url = {https://dl.acm.org/doi/10.5555/3666122.3669482},
year = {2023},
publisher = {Curran Associates Inc.},
address = {Red Hook, NY, USA},
booktitle = {Proceedings of the 37th International Conference on Neural Information Processing Systems},
articleno = {3360},
numpages = {15},
location = {New Orleans, LA, USA},
series = {NIPS '23}
}

@inproceedings{desai2024systemic,
  title={{Systemic Biases in Sign Language AI Research: A Deaf-Led Call to Reevaluate Research Agendas}},
  author={Desai, Aashaka and De Meulder, Maartje and Hochgesang, Julie A and Kocab, Annemarie and Lu, Alex X},
  booktitle={Proceedings of the LREC-COLING 2024 11th Workshop on the Representation and Processing of Sign Languages: Evaluation of Sign Language Resources},
  pages={54--65},
  url = {https://aclanthology.org/2024.signlang-1.6/},
  year={2024}
}

@inproceedings{desai2025toward,
author = {Desai, Aashaka and Alharbi, Rahaf and Hsueh, Stacy and Ladner, Richard E. and Mankoff, Jennifer},
title = {{Toward Language Justice: Exploring Multilingual Captioning for Accessibility}},
year = {2025},
isbn = {9798400713941},
publisher = {Association for Computing Machinery},
address = {New York, NY, USA},
url = {https://doi.org/10.1145/3706598.3713622},
doi = {10.1145/3706598.3713622},
booktitle = {Proceedings of the 2025 CHI Conference on Human Factors in Computing Systems},
articleno = {218},
numpages = {18},
location = {Yokohama, Japan},
series = {CHI '25}
}

@inproceedings{de2021good,
    title = {{Is ``good enough'' good enough? Ethical and responsible development of sign language technologies}},
    author = {De Meulder, Maartje},
    editor = {Shterionov, Dimitar},
    booktitle = {Proceedings of the 1st International Workshop on Automatic Translation for Signed and Spoken Languages (AT4SSL)},
    month = {Aug},
    year = {2021},
    address = {Virtual},
    publisher = {Association for Machine Translation in the Americas},
    url = {https://aclanthology.org/2021.mtsummit-at4ssl.2/},
    pages = {12--22},
}

@article{de2021sign,
  title={{Sign language interpreting services: A quick fix for inclusion?}},
  author={De Meulder, Maartje and Haualand, Hilde},
  journal={Translation and Interpreting Studies},
  volume={16},
  number={1},
  pages={19--40},
  year={2021},
  url = {https://doi.org/10.1075/tis.18008.dem},
  publisher={John Benjamins Publishing Company Amsterdam/Philadelphia}
}

@article{de2025deaf,
  title={{Deaf in AI: AI language technologies and the erosion of linguistic rights}},
  author={De Meulder, Maartje},
  url={https://ojs.letras.up.pt/index.php/LLLD/libraryFiles/downloadPublic/599},
  year={2025}
}

@misc{deaf-reddit,
  author    = {{r/deaf}},
  title     = {{NEW total ban on research affective immediately!}},
  url      = {https://www.reddit.com/r/deaf/comments/1i4gk9n/new_total_ban_on_research_affective_immediately/},
  note      = {Retrieved June 5, 2025}
}

@misc{dark-age,
  author    = {HandSpeak},
  title     = {Milan, Italy 1880},
  url      = {https://www.handspeak.com/learn/238/},       
  note      = {Retrieved May 20, 2025}
}

@book{de2019legal,
  title={The legal recognition of sign languages: Advocacy and outcomes around the world},
  author={De Meulder, Maartje and Murray, Joseph J and McKee, Rachel L},
  year={2019},
  url = {https://www.jstor.org/stable/jj.22730579},
  publisher={Multilingual Matters}
}

@article{esselink2024exploring,
  title={{Exploring automatic text-to-sign translation in a healthcare setting}},
  author={Esselink, Lyke and Roelofsen, Floris and Dotla{\v{c}}il, Jakub and Mende-Gillings, Shani and De Meulder, Maartje and Sijm, Nienke and Smeijers, Anika},
  journal={Universal Access in the Information Society},
  volume={23},
  number={1},
  pages={35--57},
  year={2024},
  url = {https://doi.org/10.1007/s10209-023-01042-6},
  publisher={Springer}
}

@article{fedorowicz2020deaf,
  title={{Deaf Bodies: Toward a Holistic Ethnography of Deaf People in Japan}},
  author={Fedorowicz, Steven C},
  journal={Journal of Inquiry and Research},
  volume={111},
  pages={269-286},
  year={2020},
  publisher={Kansai Gaidai University},
  url = {https://doi.org/10.18956/00007919}
}

@article{garg2025s,
  title={``It's trained by non-disabled people'': Evaluating How Image Quality Affects Product Captioning with VLMs},
  author={Garg, Kapil and Tang, Xinru and Heo, Jimin and Morgan, Dwayne R and Gergle, Darren and Sudderth, Erik B and Piper, Anne Marie},
  journal={arXiv preprint arXiv:2511.08917},
  url={https://arxiv.org/abs/2511.08917},
  year={2025}
}

@article{g2020science,
  title={Science communication for the Deaf in the pandemic period: absences and pursuit of information},
  author={G Silva, Alexandre and Batista, Tiago and Giraud, Felipe and Giraud, Andrea and Pinto-Silva, Flavio Eduardo and Barral, Julia and Nascimento Guimar{\~a}es, Juan and rumjanek, Vivian},
  journal={Journal of Science Communication},
  volume={19},
  number={5},
  pages={A05},
  year={2020},
  url={https://doi.org/10.22323/2.19050205},
  publisher={SISSA Medialab}
}

@misc{cctv_sc,
  author = {XueZhu},
  title = {{Paradox in Sign Language: 90\% of deaf people cannot understand the sign language used by sign language interpreters | in-depth report \begin{CJK}{UTF8}{gbsn}(吊诡的手语：手语翻译打的手语，九成聋人看不懂 | 深度)\end{CJK}}},
  url = {https://mp.weixin.qq.com/s/ERI05cuX9QpmpzumXCREaw},
  year = {Retrieved July, 2025}}

@inproceedings{glasser2020accessibility,
author = {Glasser, Abraham and Mande, Vaishnavi and Huenerfauth, Matt},
title = {{Accessibility for deaf and hard of hearing users: Sign language conversational user interfaces}},
year = {2020},
isbn = {9781450375443},
publisher = {Association for Computing Machinery},
address = {New York, NY, USA},
url = {https://doi.org/10.1145/3405755.3406158},
doi = {10.1145/3405755.3406158},
booktitle = {Proceedings of the 2nd Conference on Conversational User Interfaces},
articleno = {55},
numpages = {3},
location = {Bilbao, Spain},
series = {CUI '20}
}

@inproceedings{glasser2022asl,
author = {Glasser, Abraham and Minakov, Fyodor and Bragg, Danielle},
title = {{ASL Wiki: An Exploratory Interface for Crowdsourcing ASL Translations}},
year = {2022},
isbn = {9781450392587},
publisher = {Association for Computing Machinery},
address = {New York, NY, USA},
url = {https://doi.org/10.1145/3517428.3544827},
doi = {10.1145/3517428.3544827},
booktitle = {Proceedings of the 24th International ACM SIGACCESS Conference on Computers and Accessibility},
articleno = {16},
numpages = {13},
location = {Athens, Greece},
series = {ASSETS '22}
}

@article{glauert2006vanessa,
  title={{VANESSA -- A system for communication between Deaf and hearing people. }},
  author={Glauert, John RW and Elliott, Ralph and Cox, Stephen J and Tryggvason, Judy and Sheard, Mary},
  journal={Technology and disability},
  volume={18},
  number={4},
  pages={207--216},
  year={2006},
  doi = {10.3233/TAD-2006-18408},
  publisher={SAGE Publications Sage UK: London, England}
}

@book{gustason1980signing,
  title={{Signing exact english}},
  author={Gustason, Gerilee and Pfetzing, Donna and Zawolkow, Esther and Norris, Carolyn B},
  volume={3131},
  year={1980},
  url={http://intrpr.info/library/gustason-signing-exact-english-ch6.pdf},
  publisher={Modern Signs Press Los Alamitos, CA}
}

@article{goodwin2004competent,
  title={{A competent speaker who can't speak: The social life of aphasia}},
  author={Goodwin, Charles},
  journal={Journal of Linguistic Anthropology},
  volume={14},
  number={2},
  pages={151--170},
  year={2004},
  url = {https://www.jstor.org/stable/43102644},
  publisher={Wiley Online Library}
}

@misc{microsoft-sign-kinect,
  author    = {{Clayton, Steve}},
  title     = {{Sign language translator uses Kinect as a bridge between the deaf and hearing}},
  url      = {{https://blogs.microsoft.com/ai/sign-language-translator-uses-kinect-as-a-bridge-between-the-deaf-and-hearing/}},
  note      = {Retrieved July 30, 2025}
}

@misc{microsoft-sign,
  author    = {{Wu, Guobin and Tansley, Stewart and Stone, Lori}},
  title     = {{Opening new doors of communication for sign language users}},
  url      = {https://www.microsoft.com/en-us/research/video/opening-new-doors-of-communication-for-sign-language-users/},
  note      = {Retrieved July 30, 2025}
}

@misc{signapse,
  author    = {{Signapse}},
  title     = {{Signapse}},
  url      = {https://www.signapse.ai/},
  note      = {Retrieved July 30, 2025}
}

@misc{OmniBridge,
  author    = {{OmniBridge}},
  title     = {{OmniBridge}},
  url      = {https://omnibridge.ai/},
  note      = {Retrieved July 30, 2025}
}

@misc{SignForDeaf,
  author    = {{SignForDeaf}},
  title     = {{SignForDeaf}},
  url      = {https://www.signfordeaf.com/},
  note      = {Retrieved July 30, 2025}
}

@misc{sign-glove-ucla,
  author    = {{Chin, Matthew}},
  title     = {{Wearable-tech glove translates sign language into speech in real time}},
  url      = {https://newsroom.ucla.edu/releases/glove-translates-sign-language-to-speech},
  note      = {Retrieved August 3, 2025}
}

@misc{sign-glove,
  author    = {{NBC News}},
  title     = {{College Students Win \$10,000 Prize for Gloves that Translate Sign Language}},
  url      = {https://www.nbcnews.com/feature/college-game-plan/college-students-win-10-000-prize-gloves-translate-sign-language-n577636},
  note      = {Retrieved August 23, 2025}
}

@misc{glove_reject,
  author    = {Erard, Michael},
  title     = {{Why Sign-Language Gloves Don't Help Deaf People}},
  url      = {https://www.theatlantic.com/technology/archive/2017/11/why-sign-language-gloves-dont-help-deaf-people/545441/},
  note      = {Retrieved June 3, 2025}}

@misc{google-gemma,
  author    = {Google},
  title     = {{Developer keynote}},
  url      = {https://io.google/2025/explore/developer-keynote-1},
year = {2025},
month ={May},
  note      = {Retrieved June 3, 2025}}

@article{henner2023unsettling,
  title={{Unsettling languages, unruly bodyminds: A crip linguistics manifesto}},
  author={Henner, Jon and Robinson, Octavian},
  journal={Journal of Critical Study of Communication and Disability},
  volume={1},
  number={1},
  pages={7--37},
  url = {https://doi.org/10.48516/jcscd_2023vol1iss1.4},
  year={2023}
}

@article{hodge2023deaf,
  title={{Deaf signing diversity and signed language translations}},
  author={Hodge, Gabrielle and Goswell, Della},
  journal={Applied Linguistics Review},
  volume={14},
  number={5},
  pages={1045--1083},
  year={2023},
  url = {https://doi.org/10.1515/applirev-2020-0034},
  publisher={De Gruyter}
}

@article{hou2020signs,
  title={{Who signs? Language ideologies about deaf and hearing child signers in one family in Mexico}},
  author={Hou, Lynn},
  journal={Sign Language Studies},
  volume={20},
  number={4},
  pages={664--690},
  year={2020},
  url = {https://www.jstor.org/stable/10.2307/26984291},
  publisher={Gallaudet University Press}
}

@phdthesis{hou2016making,
  title={``Making hands'': family sign languages in the San Juan Quiahije community},
  author={Hou, Lynn Yong-Shi},
  url = {https://repositories.lib.utexas.edu/items/9023a63d-33e7-4e82-bd76-a3ad7027e82f},
  year={2016}
}

@article{harrelson2019deaf,
  title={{Deaf people with ``no language'': Mobility and flexible accumulation in languaging practices of deaf people in Cambodia}},
  author={Harrelson, Erin Moriarty},
  journal={Applied Linguistics Review},
  volume={10},
  number={1},
  pages={55--72},
  year={2019},
  url = {https://doi.org/10.1515/applirev-2017-0081},
  publisher={De Gruyter Mouton}
}

@inproceedings{huenerfauth2014release,
  author    = {Huenerfauth, Matt and Kacorri, Hernisa},
  title     = {{Release of Experimental stimuli and questions for evaluating facial expressions in animations of {American} {Sign} {Language}}},
  pages     = {71--76},
  editor    = {Crasborn, Onno and Efthimiou, Eleni and Fotinea, Stavroula-Evita and Hanke, Thomas and Hochgesang, Julie A. and Kristoffersen, Jette and Mesch, Johanna},
  booktitle = {Proceedings of the {LREC2014} 6th Workshop on the Representation and Processing of Sign Languages: Beyond the Manual Channel},
  maintitle = {9th International Conference on Language Resources and Evaluation ({LREC} 2014)},
  publisher = {{European Language Resources Association (ELRA)}},
  address   = {Reykjavik, Iceland},
  day       = {31},
  month     = may,
  year      = {2014},
  language  = {english},
  url       = {https://www.sign-lang.uni-hamburg.de/lrec/pub/14010.pdf}
}

@article{hauser2000code,
  title={{Code switching: American Sign Language and cued English}},
  author={Hauser, Peter},
  url={https://repository.rit.edu/article/325/},
  year={2000}
}

@article{hin2022translanguaging,
  title={{Translanguaging in Hong Kong Deaf Signers: Translating Meaning from Written Chinese}},
  author={Hin, Chan Yi and Lam, Anita Yu On and Leung, Aaron Wong Yiu},
  journal={Sign Language Studies},
  volume={22},
  number={3},
  pages={430--483},
  year={2022},
  url = {https://www.jstor.org/stable/27186994},
  publisher={Gallaudet University Press}
}

@article{holmstrom2018deaf,
  title={{Deaf lecturers' translanguaging in a higher education setting. A multimodal multilingual perspective}},
  author={Holmstr{\"o}m, Ingela and Sch{\"o}nstr{\"o}m, Krister},
  journal={Applied Linguistics Review},
  volume={9},
  number={1},
  pages={90--111},
  year={2018},
  url = {https://doi.org/10.1515/applirev-2017-0078},
  publisher={De Gruyter Mouton}
}

@inproceedings{inan2025signalignlm,
    title = {{{S}ign{A}lign{LM}: Integrating Multimodal Sign Language Processing into Large Language Models}},
    author = "Inan, Mert  and
      Sicilia, Anthony  and
      Alikhani, Malihe",
    editor = "Che, Wanxiang  and
      Nabende, Joyce  and
      Shutova, Ekaterina  and
      Pilehvar, Mohammad Taher",
    booktitle = "Findings of the Association for Computational Linguistics: ACL 2025",
    month = jul,
    year = "2025",
    address = "Vienna, Austria",
    publisher = "Association for Computational Linguistics",
    url = "https://aclanthology.org/2025.findings-acl.190/",
    doi = "10.18653/v1/2025.findings-acl.190",
    pages = "3691--3706",
    ISBN = "979-8-89176-256-5",}

@misc{brace-asl,
  author    = {{Brace, Aaron}},
  title     = {{Hearing Interpreters: The Danger of Being the Public Face of ASL}},
url      = {https://streetleverage.com/2014/11/hearing-interpreters-the-danger-of-being-the-public-face-of-asl/},
note ={Retrieved August, 2025}}

@article{interpretation_tv,
  title={{Survey of Sign Language Use in China \begin{CJK}{UTF8}{gbsn}(我国手语使用状况的调查研究)\end{CJK}}},
  author={Liu, Yanhong and Gu, Dingqian and Cheng, Li and Wei, Dan},
  journal={Applied Linguistics},
  number={2},
  pages={35--41},
  url = {https://qikan.cqvip.com/Qikan/Article/Detail?id=45810462&from=Qikan_Article_Detail},
  year={2013}
}

@article{jones2021nothing,
  title={Nothing about us without us: Deaf education and sign language access in China},
  author={Jones, Gabrielle A and Ni, Dawei and Wang, Wei},
  journal={Deafness \& Education International},
  volume={23},
  number={3},
  pages={179--200},
  year={2021},
  url={https://doi.org/10.1080/14643154.2021.1885576},
  publisher={Taylor \& Francis}
}

@misc{japanese-yamauchi,
  author = {{Yamauchi, Kazuhiro}},
  title = {\begin{CJK}{UTF8}{min}Japanese and Japanese Sign Language -- Toward a History of Conflict and Symbiosis -- (日本語と日本手話 ー 相克の歴史と共生に向けてー)\end{CJK}},
  howpublished = {\url{https://www.sangiin.go.jp/japanese/annai/chousa/rippou_chousa/backnumber/2017pdf/20170301101ss.pdf}},
  year = {Retrieved November, 2025}}

@article{johnston2010sampling,
  title={Sampling hard-to-reach populations with respondent driven sampling},
  author={Johnston, Lisa G and Sabin, Keith},
  journal={Methodological innovations online},
  volume={5},
  number={2},
  pages={38--48},
  year={2010},
  url = {https://doi.org/10.4256/mio.2010.0017 },
  publisher={SAGE Publications Sage UK: London, England}
}

@article{krawczyk2024ethics,
  title={Ethics of research engagement with Deaf people. A qualitative evidence synthesis},
  author={Krawczyk, Tomasz and Piasecki, Jan and Wasylewski, Mateusz and Waligora, Marcin},
  journal={Journal of Deaf Studies and Deaf Education},
  volume={29},
  number={4},
  pages={443--455},
  year={2024},
  url={https://doi.org/10.1093/jdsade/enae024},
  publisher={Oxford University Press}
}

@inproceedings{kamikubo2025exploring,
author = {Kamikubo, Rie and Glasser, Abraham and Lu, Alex X and Daum\'{e} III, Hal and Kacorri, Hernisa and Bragg, Danielle},
title = {Exploring Collaboration to Center the Deaf Community in Sign Language AI},
year = {2025},
isbn = {9798400706769},
publisher = {Association for Computing Machinery},
address = {New York, NY, USA},
url = {https://doi.org/10.1145/3663547.3746390},
doi = {10.1145/3663547.3746390},
booktitle = {Proceedings of the 27th International ACM SIGACCESS Conference on Computers and Accessibility},
articleno = {60},
numpages = {18},
location = {Denver, CO, USA},
series = {ASSETS '25}
}

@article{kusters2019language,
author={Kusters, Annelies and De Meulder, Maartje}, year={2019}, 
title={{Language Portraits: Investigating Embodied Multilingual and Multimodal Repertoires}}, 
volume={20}, 
url={https://www.qualitative-research.net/index.php/fqs/article/view/3239}, 
DOI={10.17169/fqs-20.3.3239}, 
number={3}, 
journal={Forum Qualitative Sozialforschung / Forum: Qualitative Social Research}, 
month={Sep}
}

@book{kusters2015deaf,
  title={{Deaf space in Adamorobe: An ethnographic study of a village in Ghana}},
  author={Kusters, Annelies},
  year={2015},
  url = {https://doi.org/10.2307/j.ctv2rh28wp},
  publisher={Gallaudet University Press}
}

@article{kusters2021emergence,
  title={Emergence and evolutions: Introducing sign language sociolinguistics},
  author={Kusters, Annelies and Lucas, Ceil},
  journal={Sign Language Studies},
  volume={22},
  number={2},
  pages={320--342},
  year={2021},
  url = {https://doi.org/10.1111/josl.12522},
  publisher={Gallaudet University Press}
}

@inproceedings{kezar2023sem,
author = {Kezar, Lee and Thomason, Jesse and Caselli, Naomi and Sehyr, Zed and Pontecorvo, Elana},
title = {{The Sem-Lex Benchmark: Modeling ASL Signs and their Phonemes}},
year = {2023},
isbn = {9798400702204},
publisher = {Association for Computing Machinery},
address = {New York, NY, USA},
url = {https://doi.org/10.1145/3597638.3608408},
doi = {10.1145/3597638.3608408},
booktitle = {Proceedings of the 25th International ACM SIGACCESS Conference on Computers and Accessibility},
articleno = {34},
numpages = {10},
location = {New York, NY, USA},
series = {ASSETS '23}
}

@inproceedings{kosa2025exploring,
author = {Kosa, Ben and Desai, Aashaka and Lu, Alex X and Ladner, Richard E. and Bragg, Danielle},
title = {{Exploring Reduced Feature Sets for American Sign Language Dictionaries}},
year = {2025},
isbn = {9798400713941},
publisher = {Association for Computing Machinery},
address = {New York, NY, USA},
url = {https://doi.org/10.1145/3706598.3714118},
doi = {10.1145/3706598.3714118},
booktitle = {Proceedings of the 2025 CHI Conference on Human Factors in Computing Systems},
articleno = {797},
numpages = {14},
location = {Yokohama, Japan},
series = {CHI '25}
}

@article{kusters2020sign,
  title={{Sign language ideologies: Practices and politics}},
  author={Kusters, Annelies and Green, Mara and Moriarty, Erin and Snoddon, Kristin},
  journal={Sign language ideologies in practice},
  pages={3--22},
  year={2020},
  url = {https://doi.org/10.1515/9781501510090-001},
  publisher={De Gruyter Mouton and Ishara Press}
}

@article{kusters2017innovations,
  title={{Innovations in deaf studies: Critically mapping the field}},
  author={Kusters, Annelies and De Meulder, Maartje and O'Brien, Dai and others},
  journal={Innovations in deaf studies: The role of deaf scholars},
  volume={12},
  pages={1--53},
  year={2017},
  ISBN = {9780190612184},
  publisher={Oxford University Press Oxford}
}

@article{kusters2017beyond,
  title={{Beyond languages, beyond modalities: Transforming the study of semiotic repertoires}},
  author={Kusters, Annelies and Spotti, Massimiliano and Swanwick, Ruth and Tapio, Elina},
  journal={International Journal of multilingualism},
  volume={14},
  number={3},
  pages={219--232},
  year={2017},
  url={https://doi.org/10.1080/14790718.2017.1321651},  publisher={Taylor \& Francis}
}

@article{lin2021translation,
  title={{Translation or creation? A case study of signed Chinese poetry from the perspective of multimodality theory}},
  author={Lin, Hao},
  journal={The Journal of Specialised Translation},
  number={35},
  pages={209--230},
  url = {https://doi.org/10.26034/cm.jostrans.2021.125},
  year={2021}
}

@misc{classifiers-asl,
  author    = {ASL American Sign Language},
  title     = {{``Classifiers'' American Sign Language (ASL)}},
url      = {https://www.lifeprint.com/asl101/pages-signs/classifiers/classifiers-main.htm}}

@misc{vv-deaf,
  author    = {Connecticut Deaf Theatre},
  title     = {{Visual Vernacular: A Global Phenomenon}},
url      = {https://www.conndeaftheatre.org/posts/visual-vernacular-a-global-phenomenon},
  note      = {Retrived November , 2025}
}

@inproceedings{liu2023visual,
author = {Liu, Xingyu ``Bruce'' and Kirilyuk, Vladimir and Yuan, Xiuxiu and Olwal, Alex and Chi, Peggy and Chen, Xiang ``Anthony'' and Du, Ruofei},
title = {{Visual Captions: Augmenting Verbal Communication with On-the-fly Visuals}},
year = {2023},
isbn = {9781450394215},
publisher = {Association for Computing Machinery},
address = {New York, NY, USA},
url = {https://doi.org/10.1145/3544548.3581566},
doi = {10.1145/3544548.3581566},
booktitle = {Proceedings of the 2023 CHI Conference on Human Factors in Computing Systems},
articleno = {108},
numpages = {20},
location = {Hamburg, Germany},
series = {CHI '23}
}

@article{love2017languaging,
  title={{On languaging and languages}},
  author={Love, Nigel},
  journal={Language Sciences},
  volume={61},
  pages={113--147},
  year={2017},
  url = {https://doi.org/10.1016/j.langsci.2017.04.001},
  publisher={Elsevier}
}

@article{lu2025sound,
  title={{The sound of silence: Chinese Deaf creators' self-presentation, labour practices and the visibility paradox on Douyin}},
  author={Lu, Pengyun and Guo, Zhiling},
  journal={New Media \& Society},
  pages={14614448251338497},
  year={2025},
  url = {https://doi.org/10.1177/14614448251338497},
  publisher={SAGE Publications Sage UK: London, England}
}

@article{lytle2005deaf,
  title={{Deaf education in China: History, current issues, and emerging deaf voices}},
  author={Lytle, Richard R and Johnson, Kathryn E and Hui, Yang Jun},
  journal={American annals of the deaf},
  volume={150},
  number={5},
  pages={457--469},
  year={2005},
  url = {https://dx.doi.org/10.1353/aad.2006.0009},
  publisher={JSTOR}
}

@book{metzger1999sign,
  title={Sign language interpreting: Deconstructing the myth of neutrality},
  author={Metzger, Melanie},
  year={1999},
  publisher={Gallaudet University Press}
}

@phdthesis{ma2020study,
  title={A Study of Lexical Variation, Comprehension and Language Attitudes in Deaf Users of Chinese Sign Language (CSL) from Beijing and Shanghai},
  author={Ma, Yunyi},
  year={2020},
  url={https://discovery.ucl.ac.uk/id/eprint/10096564},
  school={UCL (University College London)}
}

@misc{mouthing-asl,
  author    = {ASL University},
  title     = {{ASL Linguistics: Mouthing in ASL / Mouth Morphemes}},
url      = {https://www.lifeprint.com/asl101/pages-layout/mouthinginasl.htm},
note = {Retrieved November 6, 2025}}

@article{mack2020social,
author = {Mack, Kelly and Bragg, Danielle and Morris, Meredith Ringel and Bos, Maarten W. and Albi, Isabelle and Monroy-Hern\'{a}ndez, Andr\'{e}s},
title = {Social App Accessibility for Deaf Signers},
year = {2020},
issue_date = {October 2020},
publisher = {Association for Computing Machinery},
address = {New York, NY, USA},
volume = {4},
number = {CSCW2},
url = {https://doi.org/10.1145/3415196},
doi = {10.1145/3415196},
journal = {Proc. ACM Hum.-Comput. Interact.},
month = oct,
articleno = {125},
numpages = {31}}

@article{mitchell2023many,
  title={{How many people use sign language? A national health survey-based estimate}},
  author={Mitchell, Ross E and Young, Travas A},
  journal={Journal of Deaf Studies and Deaf Education},
  volume={28},
  number={1},
  pages={1--6},
  year={2023},
  url = {https://doi.org/10.1093/deafed/enac031},
  publisher={Oxford University Press}
}

@misc{asl-great,
  author    = {Okrent, Arika},
  title     = {{Why Great Sign Language Interpreters Are So Animated}},
  url      = {https://www.theatlantic.com/health/archive/2012/11/why-great-sign-language-interpreters-are-so-animated/264459/},
  note      = {Retrieved August 3, 2025}}

@article{maler2013songs,
  title={{Songs for hands: Analyzing interactions of sign language and music}},
  author={Maler, Anabel},
  journal={Music theory online},
  volume={19},
  number={1},
  url = {https://doi.org/10.30535/mto.19.1.4},
  year={2013}
}

@misc{nvdia-asl,
  author    = {Boone, Michael},
  title     = {{It's a Sign: AI Platform for Teaching American Sign Language Aims to Bridge Communication Gaps}},
  url      = {https://blogs.nvidia.com/blog/ai-sign-language/},
  note      = {Retrieved June 3, 2025}}

@misc{ndc-deaf,
  author    = {{National Deaf Center}},
  title     = {{Communicating With Deaf Individuals}},
  url      = {https://nationaldeafcenter.org/resource-items/communicating-deaf-people/},
year = {2025},
month = {May},
day = {12},
  note      = {Retrieved July, 2025}}

@book{nakamura2006deaf,
  title={{Deaf in Japan: Signing and the politics of identity}},
  author={Nakamura, Karen},
  year={2006},
  publisher={Cornell University Press}
}

@incollection{palfreyman2019variation,
  title={{Variation in Indonesian sign language}},
  author={Palfreyman, Nick},
  year={2019},
  publisher={De Gruyter Mouton}
}

@book{padden2009inside,
  title={{Inside deaf culture}},
  author={Padden, Carol and Humphries, Tom},
  year={2009},
  publisher={Harvard University Press}
}

@article{pennycook2017translanguaging,
  title={{Translanguaging and semiotic assemblages}},
  author={Pennycook, Alastair},
  journal={International Journal of Multilingualism},
  volume={14},
  number={3},
  pages={269--282},
  year={2017},
  url = {https://doi.org/10.1080/14790718.2017.1315810},
  publisher={Taylor \& Francis}
}

@article{prietch2022systematic,
author = {Prietch, Soraia and S\'{a}nchez, J. Alfredo and Guerrero, Josefina},
title = {{A Systematic Review of User Studies as a Basis for the Design of Systems for Automatic Sign Language Processing}},
year = {2022},
issue_date = {December 2022},
publisher = {Association for Computing Machinery},
address = {New York, NY, USA},
volume = {15},
number = {4},
issn = {1936-7228},
url = {https://doi.org/10.1145/3563395},
doi = {10.1145/3563395},
journal = {ACM Trans. Access. Comput.},
month = nov,
articleno = {36},
numpages = {33}
}

@article{pollard2009adapting,
  title={{Adapting health education material for deaf audiences.}},
  author={Pollard Jr, Robert Q and Dean, Robyn K and O'Hearn, Amanda and Haynes, Sharon L},
  journal={Rehabilitation psychology},
  volume={54},
  number={2},
  pages={232},
  year={2009},
  url = {https://doi.org/10.1037/a0015772},
  publisher={American Psychological Association}
}

@book{padden1988deaf,
  title={{Deaf in America: Voices from a culture}},
  author={Padden, Carol and Humphries, Tom},
  year={1988},
  publisher={Harvard University Press}
}

@article{ren2024influence,
author = {Ren, Tianyu and Yao, Dengfeng and Yang, Chaoran and Kang, Xinchen},
title = {The Influence of Chinese Characters on Chinese Sign Language},
year = {2024},
issue_date = {January 2024},
publisher = {Association for Computing Machinery},
address = {New York, NY, USA},
volume = {23},
number = {1},
issn = {2375-4699},
url = {https://doi.org/10.1145/3591465},
doi = {10.1145/3591465},
journal = {ACM Trans. Asian Low-Resour. Lang. Inf. Process.},
month = jan,
articleno = {6},
numpages = {31}
}

@article{ramati2018uniform,
  title={{Uniform multilingualism: A media genealogy of Google Translate}},
  author={Ramati, Ido and Pinchevski, Amit},
  journal={New media \& society},
  volume={20},
  number={7},
  pages={2550--2565},
  year={2018},
  url={https://doi.org/10.1177/1461444817726951},
  publisher={SAGE Publications Sage UK: London, England}
}

@article{schmitz2021deaf,
  title={Deaf-Queer Signing in Process: A Qualitative Sociolinguistic Study of ``Queering Deafhood,'' ``Deafing Queerhood,'' and ``Queer Sign Language Style''},
  author={Schmitz, Jona},
  journal={Sign Language Studies},
  volume={22},
  number={1},
  pages={42--74},
  year={2021},
  url={https://doi.org/10.1353/sls.2021.0014},
  publisher={Gallaudet University Press}
}

@article{stokoe2005sign,
  title={Sign language structure: An outline of the visual communication systems of the American deaf},
  author={Stokoe Jr, William C},
  journal={Journal of deaf studies and deaf education},
  volume={10},
  number={1},
  pages={3--37},
  year={2005},
  url={https://doi.org/10.1093/deafed/eni001},
  publisher={Oxford University Press}
}

@article{skinner2018interpreting,
  title={{Interpreting via video link: Mapping of the field}},
  author={Skinner, Robert and Napier, Jemina and Braun, Sabine},
  journal={Here or there: Research on interpreting via video link},
  url = {https://doi.org/10.2307/j.ctv2rh2bs3},
  pages={11--35},
  year={2018},
  publisher={Gallaudet University Press Washington}
}

@inproceedings{suchanek2025participation,
author = {Suchanek, Oliver and Meissner, Janis Lena and Angelini, Robin and Spiel, Katta},
title = {{From Participation to Solidarity: A Case Study on Access of Maker Spaces from Deaf and Hearing Perspectives}},
year = {2025},
isbn = {9798400713941},
publisher = {Association for Computing Machinery},
address = {New York, NY, USA},
url = {https://doi.org/10.1145/3706598.3713202},
doi = {10.1145/3706598.3713202},
booktitle = {Proceedings of the 2025 CHI Conference on Human Factors in Computing Systems},
articleno = {292},
numpages = {15},
location = {Yokohama, Japan},
series = {CHI '25}
}

@misc{starbucks-japan,
  author    = {McGee, Oona},
  title     = {{Starbucks Japan opens first sign-language store in Tokyo}},
url      = {https://soranews24.com/2020/06/29/starbucks-japan-opens-first-sign-language-store-in-tokyo/},
  note      = {Retrieved August 27, 2025}}

@misc{starbucks,
  author    = {Starbucks},
  title     = {{Starbucks Coffee Company}},
url      = {https://www.starbucks.com/},
  note      = {Retrieved August 27, 2025}}

@article{sturman1994survey,
  title={{A survey of glove-based input}},
  author={Sturman, David J and Zeltzer, David},
  journal={IEEE Computer graphics and Applications},
  volume={14},
  number={1},
  pages={30--39},
  year={1994},
  doi={10.1109/38.250916},
  publisher={IEEE}
}

@INPROCEEDINGS{starner1995real,
  author={Starner, T. and Pentland, A.},
  booktitle={Proceedings of International Symposium on Computer Vision - ISCV}, 
  title={{Real-time American Sign Language recognition from video using hidden Markov models}}, 
  year={1995},
  volume={},
  number={},
  pages={265-270},
  doi={10.1109/ISCV.1995.477012}}

@article{starner2002real,
  title={{Real-time american sign language recognition using desk and wearable computer based video}},
  author={Starner, Thad and Weaver, Joshua and Pentland, Alex},
  journal={IEEE Transactions on pattern analysis and machine intelligence},
  volume={20},
  number={12},
  pages={1371--1375},
  year={2002},
  doi={10.1109/34.735811},
  publisher={IEEE}
}

@inproceedings{tang2026disability,
author = {Tang, Xinru and Li, Jingjin and Wu, Shaomei},
title = {Disability-First AI Dataset Annotation: Co-designing Stuttered Speech Annotation Guidelines with People Who Stutter},
year = {2026},
isbn = {979-8-4007-2278-3/2026/04},
publisher = {Association for Computing Machinery},
address = {New York, NY, USA},
url = {https://doi.org/10.1145/3772318.3790405},
doi = {10.1145/3772318.3790405},
booktitle = {Proceedings of the 2026 CHI Conference on Human Factors in Computing Systems},
numpages = {22},
location = {Barcelona, Spain},
series = {CHI '26}
}

@article{tang2025beyond,
author = {Tang, Xinru and Lima, Gabriel and Jiang, Li, Jiang and Simko, Lucy and Zou, Yixin},
title = {Beyond ``Vulnerable Populations'': A Unified Understanding of Vulnerability From A Socio-Ecological Perspective},
year = {2025},
issue_date = {May 2025},
publisher = {Association for Computing Machinery},
address = {New York, NY, USA},
volume = {9},
number = {2},
url = {https://doi.org/10.1145/3710935},
doi = {10.1145/3710935},
journal = {Proc. ACM Hum.-Comput. Interact.},
month = may,
articleno = {CSCW037},
numpages = {30}
}

@inproceedings{tran2026toward,
  title={Toward Inclusive Security and Privacy for Deaf and Hard-of-Hearing People: A Community-Based Interview Study},
  author={Tran, Mindy and Tang, Xinru and Hutchinson, Adryana and Aviv, Adam J and Zou, Yixin},
  booktitle={2026 IEEE Symposium on Security and Privacy (S\&P)},
  year={2026},
  organization={IEEE}
}

@misc{daily-moth,
  author    = {The Daily Moth},
  title     = {{The Daily Moth}},
url      = {https://members.dailymoth.com/},
note = {Retrieved November 6, 2025}}

@inproceedings{tran2023us,
author = {Tran, Nina and Ladner, Richard E. and Bragg, Danielle},
title = {{U.S. Deaf Community Perspectives on Automatic Sign Language Translation}},
year = {2023},
isbn = {9798400702204},
publisher = {Association for Computing Machinery},
address = {New York, NY, USA},
url = {https://doi.org/10.1145/3597638.3614507},
doi = {10.1145/3597638.3614507},
booktitle = {Proceedings of the 25th International ACM SIGACCESS Conference on Computers and Accessibility},
articleno = {76},
numpages = {7},
location = {New York, NY, USA},
series = {ASSETS '23}
}

@misc{trs-fcc,
  author    = {{Federal Communication Commission}},
  title     = {{Telecommunications Relay Service - TRS}},
  url      = {https://www.fcc.gov/consumers/guides/telecommunications-relay-service-trs},
year = {2022},
month ={August},
day ={16},
  note      = {Retrieved June 3, 2025}
}

@inproceedings{xinru_deaf,
author = {Tang, Xinru and Chang, Xiang and Chen, Nuoran and Ni, Yingjie (MaoMao) and LC, RAY and Tong, Xin},
title = {{Community-Driven Information Accessibility: Online Sign Language Content Creation within d/Deaf Communities}},
year = {2023},
isbn = {9781450394215},
publisher = {Association for Computing Machinery},
address = {New York, NY, USA},
url = {https://doi.org/10.1145/3544548.3581286},
doi = {10.1145/3544548.3581286},
booktitle = {Proceedings of the 2023 CHI Conference on Human Factors in Computing Systems},
articleno = {50},
numpages = {24},
location = {Hamburg, Germany},
series = {CHI '23}
}

@inproceedings{vogler2025barriers,
author = {Vogler, Christian and Glasser, Abraham and Kushalnagar, Raja and Seita, Matthew and Arroyo Chavez, Mariana and Delk, Keith and DeVries, Paige and Feanny, Molly and Thompson, Bernard and Waller, James},
title = {{Barriers to Employment: The Deaf Multimedia Authoring Tax}},
year = {2025},
isbn = {9798400718823},
publisher = {Association for Computing Machinery},
address = {New York, NY, USA},
url = {https://doi.org/10.1145/3744257.3744269},
doi = {10.1145/3744257.3744269},
booktitle = {Proceedings of the 22nd International Web for All Conference},
pages = {95-99},
numpages = {5},
location = {Sydney, Australia},
series = {W4A '25}
}

@article{wehrmeyer2015comprehension,
  title={Comprehension of television news signed language interpreters: A South African perspective},
  author={Wehrmeyer, Ella},
  journal={Interpreting},
  volume={17},
  number={2},
  pages={195--225},
  year={2015},
  url={https://doi.org/10.1075/intp.17.2.03weh},
  publisher={John Benjamins}
}

@misc{sign-avatar,
  author    = {WIRED},
  title     = {{This Startup Has Created AI-Powered Signing Avatars for the Deaf}},
  url      = {https://www.wired.com/story/silence-speaks-deaf-ai-signing/},
year = {2025},
month ={May},
day={7},
  note      = {Retrieved June 3, 2025}}

@article{wang2018accessibility,
author = {Wang, Emily Q. and Piper, Anne Marie},
title = {{Accessibility in Action: Co-Located Collaboration among Deaf and Hearing Professionals}},
year = {2018},
issue_date = {November 2018},
publisher = {Association for Computing Machinery},
address = {New York, NY, USA},
volume = {2},
number = {CSCW},
url = {https://doi.org/10.1145/3274449},
doi = {10.1145/3274449},
journal = {Proc. ACM Hum.-Comput. Interact.},
month = nov,
articleno = {180},
numpages = {25},
}

@article{wei2018translanguaging,
  title={{Translanguaging as a practical theory of language}},
  author={Wei, Li},
  journal={Applied linguistics},
  volume={39},
  number={1},
  pages={9--30},
  year={2018},
  url = {https://doi.org/10.1093/applin/amx039},
  publisher={Oxford University Press}
}

@article{xiao2015chinese,
  title={Chinese Deaf viewers' comprehension of sign language interpreting on television: An experimental study},
  author={Xiao, Xiaoyan and Chen, Xiaoyan and Palmer, Jeffrey Levi},
  journal={Interpreting},
  volume={17},
  number={1},
  pages={91--117},
  year={2015},
  url ={https://doi.org/10.1075/intp.17.1.05xia},
  publisher={John Benjamins}
}

@inproceedings{dengfeng-csl-classifiers,
  title={{Cognitive-semantic Analysis of Classifier Predicates in Chinese Sign Language \begin{CJK}{UTF8}{gbsn}(论中国手语的分类词谓语)\end{CJK}}},
  author={Yao, Dengfeng and Jiang, Minghu and Chang, Jung-hsing and Abulizi, Abudoukelimu},
  booktitle={Journal of Chinese Information Processing},
  pages={1--8},
  url = {http://jcip.cipsc.org.cn/CN/abstract/abstract2526.shtml},
  year={2018}
}

@inproceedings{yin2024asl,
    title = {{{ASL} {STEM} {W}iki: Dataset and Benchmark for Interpreting {STEM} Articles}},
    author = "Yin, Kayo  and
      Singh, Chinmay  and
      Minakov, Fyodor O  and
      Milan, Vanessa  and
      Daum{\'e} Iii, Hal  and
      Zhang, Cyril  and
      Lu, Alex Xijie  and
      Bragg, Danielle",
    editor = "Al-Onaizan, Yaser  and
      Bansal, Mohit  and
      Chen, Yun-Nung",
    booktitle = "Proceedings of the 2024 Conference on Empirical Methods in Natural Language Processing",
    month = nov,
    year = "2024",
    address = "Miami, Florida, USA",
    publisher = "Association for Computational Linguistics",
    url = "https://aclanthology.org/2024.emnlp-main.801/",
    doi = "10.18653/v1/2024.emnlp-main.801",
    pages = "14474--14490",
}

@inproceedings{yoo2025elmi,
author = {Yoo, Suhyeon and Truong, Khai N. and Kim, Young-Ho},
title = {ELMI: Interactive and Intelligent Sign Language Translation of Lyrics for Song Signing},
year = {2025},
isbn = {9798400713941},
publisher = {Association for Computing Machinery},
address = {New York, NY, USA},
url = {https://doi.org/10.1145/3706598.3713973},
doi = {10.1145/3706598.3713973},
booktitle = {Proceedings of the 2025 CHI Conference on Human Factors in Computing Systems},
articleno = {566},
numpages = {21},
location = {Yokohama, Japan},
series = {CHI '25}
}

@inproceedings{yoo2023understanding,
author = {Yoo, Suhyeon and Lin, Georgianna and Byeon, Hyeon Jeong and Hwang, Amy S. and Truong, Khai Nhut},
title = {{Understanding tensions in music accessibility through song signing for and with d/Deaf and Non-d/Deaf persons}},
year = {2023},
isbn = {9781450394215},
publisher = {Association for Computing Machinery},
address = {New York, NY, USA},
url = {https://doi.org/10.1145/3544548.3581287},
doi = {10.1145/3544548.3581287},
booktitle = {Proceedings of the 2023 CHI Conference on Human Factors in Computing Systems},
articleno = {59},
numpages = {18},
location = {Hamburg, Germany},
series = {CHI '23}
}

@article{young2019translated,
  title={{The translated deaf self, ontological (in) security and deaf culture}},
  author={Young, Alys and Napier, Jemina and Oram, Rosemary},
  journal={The translator},
  volume={25},
  number={4},
  pages={349--368},
  year={2019},
  url = {https://doi.org/10.1080/13556509.2020.1734165},
  publisher={Taylor \& Francis}
}

@article{young2016understanding,
  title={Understanding dementia: effective information access from the deaf community's perspective},
  author={Young, Alys and Ferguson-Coleman, Emma and Keady, John},
  journal={Health \& Social Care in the Community},
  volume={24},
  number={1},
  pages={39--47},
  year={2016},
  doi = {10.1111/hsc.12181},
  publisher={Wiley Online Library}
}

@inproceedings{zafrulla2011american,
author = {Zafrulla, Zahoor and Brashear, Helene and Starner, Thad and Hamilton, Harley and Presti, Peter},
title = {{American sign language recognition with the kinect}},
year = {2011},
isbn = {9781450306416},
publisher = {Association for Computing Machinery},
address = {New York, NY, USA},
url = {https://doi.org/10.1145/2070481.2070532},
doi = {10.1145/2070481.2070532},
booktitle = {Proceedings of the 13th International Conference on Multimodal Interfaces},
pages = {279-286},
numpages = {8},
location = {Alicante, Spain},
series = {ICMI '11}
}

@inproceedings{zhang2025towards,
author = {Zhang, Han and Shalev-Arkushin, Rotem and Baltatzis, Vasileios and Gillis, Connor and Laput, Gierad and Kushalnagar, Raja and Quandt, Lorna C and Findlater, Leah and Bedri, Abdelkareem and Lea, Colin},
title = {{Towards AI-driven Sign Language Generation with Non-manual Markers}},
year = {2025},
isbn = {9798400713941},
publisher = {Association for Computing Machinery},
address = {New York, NY, USA},
url = {https://doi.org/10.1145/3706598.3713855},
doi = {10.1145/3706598.3713855},
booktitle = {Proceedings of the 2025 CHI Conference on Human Factors in Computing Systems},
articleno = {278},
numpages = {26},
location = {Yokohama, Japan},
series = {CHI '25}
}
\appendix
\clearpage
\onecolumn
\section{Translation Strategies Observed Among Deaf Creators}
\label{appendix::translation}
\setcounter{table}{0} 
\begin{table}[h]
\resizebox{\textwidth}{!}{
\begin{tabular}{cllll}
\hline
\multicolumn{1}{l}{}                                                                                                   &                                                                                                 & \textbf{Modalities} & \textbf{Systems or Strategies}                                                                                      & \textbf{Examples or Purposes}                                                                                                                                                      \\ \hline
\multicolumn{1}{c|}{\multirow{8}{*}{\begin{tabular}[c]{@{}c@{}}Language\\ Systems\end{tabular}}}                       & \multicolumn{1}{l|}{Gloss}                                                                      & Text                & Gloss                                                                                                               & \begin{tabular}[c]{@{}l@{}}adding Gloss captions to \\ help non-signers recognize signs\end{tabular}                                                                               \\ \cline{2-5} 
\multicolumn{1}{c|}{}                                                                                                  & \multicolumn{1}{l|}{\multirow{3}{*}{Chinese}}                                                   & Text                & \begin{tabular}[c]{@{}l@{}}Written \\ Chinese\end{tabular}                                                          & \begin{tabular}[c]{@{}l@{}}displaying the original Chinese\\ word or adding Chinese captions\\ to ground signs without \\ standardized translations or\\ deaf accents, i.e., DHH people's\\speech may sound different from\\hearing individuals \end{tabular} \\ \cline{3-5} 
\multicolumn{1}{c|}{}                                                                                                  & \multicolumn{1}{l|}{}                                                                           & Speech              & \begin{tabular}[c]{@{}l@{}}Spoken \\ Mandarin\end{tabular}                                                           & \begin{tabular}[c]{@{}l@{}}using spoken Mandarin \\ to attract hearing audiences\end{tabular}                                                                                       \\ \cline{3-5} 
\multicolumn{1}{c|}{}                                                                                                  & \multicolumn{1}{l|}{}                                                                           & Speech              & \begin{tabular}[c]{@{}l@{}}AI-generated \\ Speech\end{tabular}                                                      & \begin{tabular}[c]{@{}l@{}}using AI-generated speech \\ to attract hearing audiences\end{tabular}                                                                                  \\ \cline{2-5} 
\multicolumn{1}{c|}{}                                                                                                  & \multicolumn{1}{l|}{\multirow{4}{*}{\begin{tabular}[c]{@{}l@{}}Signing\\ Systems\end{tabular}}} & Visual              & CSL Variants                                                                                                        & \begin{tabular}[c]{@{}l@{}}picking the most common\\ signs out of CSL variants\end{tabular}                                                                                        \\ \cline{3-5} 
\multicolumn{1}{c|}{}                                                                                                  & \multicolumn{1}{l|}{}                                                                           & Visual              & \begin{tabular}[c]{@{}l@{}}Signed Chinese\\or fingerspelling \end{tabular}                                                                                                        & \begin{tabular}[c]{@{}l@{}}fingerspelling `\textcolor{purple}{Ch-M-W-L}' to\\translate the Chinese idiom\\ \begin{CJK}{UTF8}{gbsn}`魑 (\textcolor{purple}{Chī}) 魅 (\textcolor{purple}{Mèi}) 魍 (\textcolor{purple}{Wǎng}) 魉(\textcolor{purple}{Liǎng})\end{CJK}'\end{tabular}                                       \\ \cline{3-5} 
\multicolumn{1}{c|}{}                                                                                                  & \multicolumn{1}{l|}{}                                                                           & Visual              & Classifiers                                                                                                         & \begin{tabular}[c]{@{}l@{}}using classifiers and \\ visual-spatial signing\\styles that native\\ signers most familiar with\end{tabular}                                       \\ \cline{3-5} 

\multicolumn{1}{c|}{}                                                                                                  & \multicolumn{1}{l|}{}                                                                           & Visual              & Mouthing                                                                                                            & \begin{tabular}[c]{@{}l@{}}Mouthing Chinese to help \\ viewers relate signs to \\ the original Chinese word\end{tabular}                                                           \\ \hline
\multicolumn{1}{c|}{\begin{tabular}[c]{@{}c@{}}Semiotic\\ Systems\end{tabular}}                                        & \multicolumn{1}{l|}{\begin{tabular}[c]{@{}l@{}}Visual\\ Elements\end{tabular}}                  & Visual              & Images                                                                                                              & \begin{tabular}[c]{@{}l@{}}adding illustrations or\\ visuals to explain concepts \\ like COVID-19\end{tabular}                                                                     \\ \hline
\multicolumn{1}{l|}{\multirow{3}{*}{\begin{tabular}[c]{@{}c@{}}Communication\\ Strategies \\ in General\end{tabular}}} & \multicolumn{1}{l|}{\begin{tabular}[c]{@{}l@{}}Adding\\ Narratives\end{tabular}}                & /                   & \begin{tabular}[c]{@{}l@{}}storytelling, \\ role-playing, \\ adding examples,\\ making analogies, etc.\end{tabular} & \begin{tabular}[c]{@{}l@{}}situating concepts in narrations,\\ e.g., role-playing mental\\ health consultations\end{tabular}                                                       \\ \cline{2-5} 
\multicolumn{1}{l|}{}                                                                                                  & \multicolumn{1}{l|}{\begin{tabular}[c]{@{}l@{}}Reducing\\ Ambiguity\end{tabular}}               & /                   & \begin{tabular}[c]{@{}l@{}}Setting\\ Contexts\end{tabular}                                                          & \begin{tabular}[c]{@{}l@{}}explaining it is an idiom\\ before translating Chinese\\ idioms\end{tabular}                                                                            \\ \cline{2-5} 
\multicolumn{1}{l|}{}                                                                                                  & \multicolumn{1}{l|}{Emphasis}                                                                   & /                   & Repetitions                                                                                                         & \begin{tabular}[c]{@{}l@{}}repeating signs when\\ introducing uncommon words\end{tabular}                                                                                          \\ \hline
\end{tabular}
}
\caption{Translation strategies participants mentioned. This list is not intended to be an exhaustive reflection of their translation but to show the multilingual, multimodal, and multicultural nature of their work.}
\label{table::examples}
\end{table}

\clearpage
\section{Video Interfaces of Platforms Participants Used}
\label{appendix::interfaces}
\setcounter{figure}{0} 
\BeginAccSupp{method=pdfstringdef,unicode,Alt={A figure showing screenshots from four different video platforms used by participants. The platforms are labeled Bilibili, WeChat Article, Kuaishou, Douyin, and Xiaohongshu. The Bilibili screenshot shows a landscape video player, typical of long-form content. The WeChat Article screenshot displays a written article with an embedded video. The video is labeled "Attached Videos." The Kuaishou, Douyin, and Xiaohongshu screenshots each show a portrait-oriented, full-screen video, characteristic of short-form video sharing.}}
\begin{figure}[h]
\centering
\includegraphics[width=14cm]{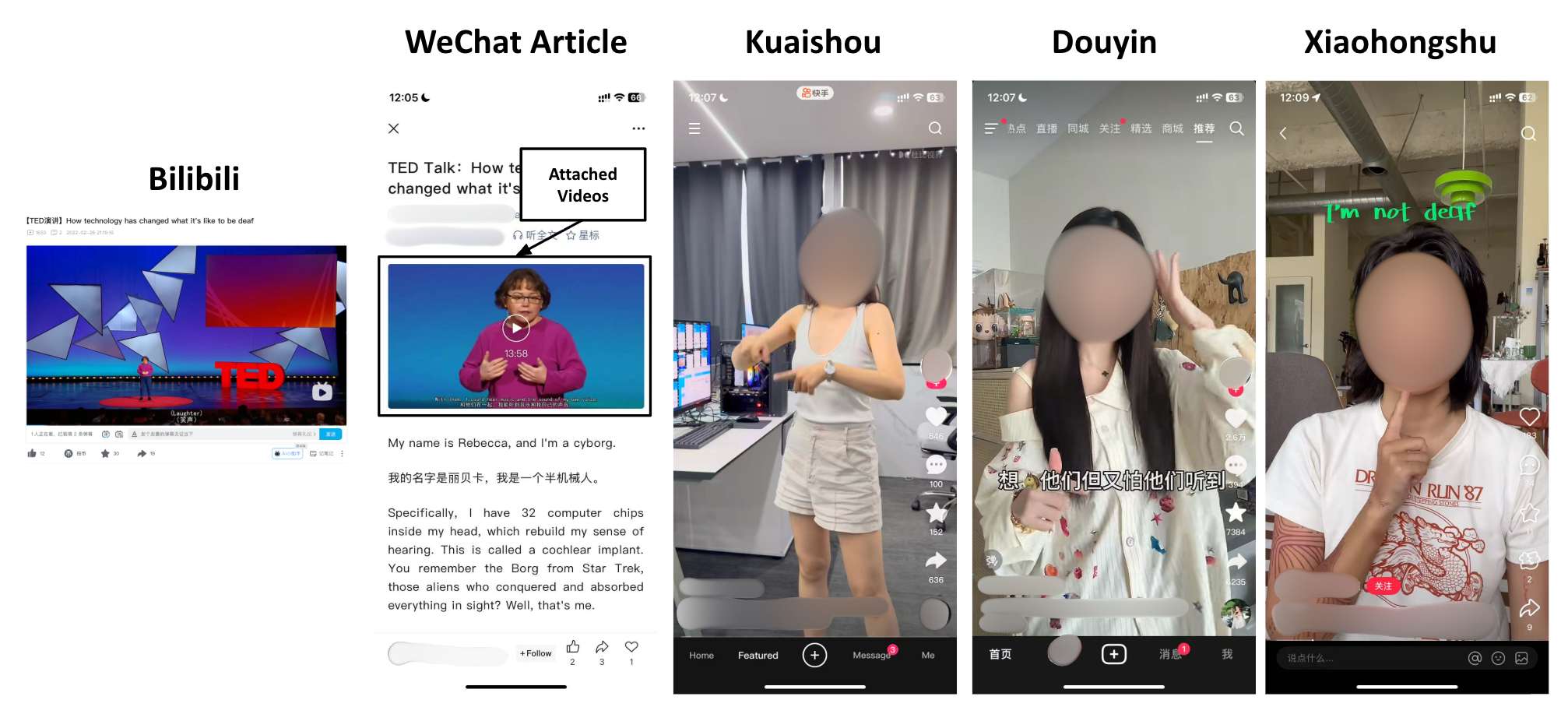}
\caption{{Screenshots of the video interfaces from platforms used by participants. Bilibili is mainly designed for long-form video content, similar to YouTube. WeChat Articles primarily host written content but allow embedded videos. Kuaishou, Douyin, and Xiaohongshu are designed for short video sharing. All platforms include common content sharing features such as ``Like'' and ``Forward.''}}
\end{figure}
\EndAccSupp{}
\end{document}